\documentclass[prd,reprint, nofootinbib,amsmath,amssymb,amsfonts,aps,
superscriptaddress,eqsecnum, floatfix, longbibliography]{revtex4-1}
\usepackage[colorlinks=true,allcolors=blue]{hyperref}

\begin{document}

\title{Center-of-mass angular momentum and memory effect in
asymptotically flat spacetimes}
\author{David A.~Nichols}
\email{d.a.nichols@uva.nl}
\affiliation{Gravitation Astroparticle Physics Amsterdam (GRAPPA), 
University of Amsterdam, Science Park, P.O.~Box 94485, 1090 GL Amsterdam, 
The Netherlands}
\affiliation{Department of Astrophysics/IMAPP, Faculty of Science, Radboud
University, P.O.~Box 9010, 6500 GL Nijmegen, The Netherlands}

\begin{abstract}
Gravitational-wave (GW) memory effects are constant changes in the GW strain 
and its time integrals, which are closely connected to changes in the charges 
that characterize asymptotically flat spacetimes.
The first GW memory effect discovered was a lasting change in the GW strain.
It can occur when GWs or massless fields carry away 4-momentum from an 
isolated source.
Subsequently, it was shown that fluxes of intrinsic angular momentum can 
generate a new type of memory effect called the spin memory, which is an 
enduring change in a portion of the time integral of the GW strain.
In this paper, we note that there is another new type of memory effect.
We call it the \textit{center-of-mass (CM) memory effect}, because it is 
related to changes in the CM part of the angular momentum of a spacetime.
We first examine a few properties of the CM angular momentum.
Specifically, we describe how it transforms under the supertranslation symmetry
transformations of the Bondi-Metzner-Sachs group, and we
compute a new expression for the flux of CM angular momentum carried by GWs
in terms of a set of radiative multipole moments of the GW strain.
We then turn to the CM memory effect.
The CM memory effect appears in a quantity which has the units of the time 
integral of the GW strain.
We define the effect in asymptotically flat spacetimes that start in a 
stationary state, radiate, and settle to a different stationary state. 
We show that it is invariant under infinitesimal supertranslation symmetries
in this context.
To determine the magnitude of the flux of CM angular momentum and the CM memory
effect, we compute these quantities for nonspinning, quasicircular compact 
binaries in the post-Newtonian approximation.
The CM memory effect arises from terms in the gravitational waveform 
for such binaries beginning at third and fourth post-Newtonian order for 
unequal- and equal-mass binaries, respectively.
Finally, we estimate the amplitude of the CM memory effect for these binaries.
We anticipate that it will be unlikely for current or upcoming GW
detectors to measure the effect.
\end{abstract}

\maketitle

\tableofcontents

\section{\label{sec:intro} Introduction}

Far from an isolated gravitating source, spacetime can be described as
asymptotically flat if it satisfies the conditions set forth by Bondi
\textit{et al.}~\cite{Bondi1962} and Sachs~\cite{Sachs1962a,Sachs1962b}
(see also, e.g., the review~\cite{Madler:2016xju}).
These spacetimes encompass the asymptotic region of a wide range of 
interesting astrophysical systems.
The gravitational waveforms used for detecting the five binary-black-hole 
mergers by the LIGO-Virgo 
Collaboration~\cite{Abbott:2016blz,TheLIGOScientific:2016pea,Abbott:2017vtc,
Abbott:2017gyy,Abbott:2017oio} and the one binary-neutron-star 
merger~\cite{TheLIGOScientific:2017qsa}, for example, are determined from
numerical simulations with asymptotically flat boundary conditions.
The symmetry group of asymptotically flat spacetimes is the  
Bondi-Metzner-Sachs (BMS) group, which consists of the Lorentz transformations 
and an infinite-dimensional, abelian group called the supertranslations.
The supertranslations include the four spacetime translations, but they predate
and are not related to supersymmetry.
Related to all the infinitesimal BMS symmetries are corresponding charges 
(see, e.g.,~\cite{Wald:1999wa}).
For the Lorentz symmetries, the conjugate charges are the angular momenta 
[which can be split into the spin and center-of-mass (CM) parts]; for the 
supertranslations, the charges are called supermomenta (by analogy with how 
the charges related to the four spacetime translations are called 4-momenta).

More recently, the symmetries of asymptotically flat spacetimes have been 
reexamined, and larger symmetry algebras than the BMS algebra have been 
proposed (see, 
e.g.,~\cite{Barnich2009,Barnich2010,Campiglia:2014yka,Campiglia:2015yka}).
The extensions of the BMS algebra involve enlargements of the Lorentz 
part of the algebra, and the conjugate charges can be thought of as 
generalizations of relativistic angular momentum.
These charges were called superspin and super CM in~\cite{Flanagan:2015pxa}, 
by analogy with nomenclatures used to describe the Lorentz charges and the 
supermomentum charges.
Collectively, we will call these charges ``super angular momentum''
(though they have also been called super-rotation charges~\cite{Barnich:2011mi} 
after the name given to the extended symmetry vector fields).
Thus, there may be an infinite number of additional charges that 
characterize an asymptotically flat spacetime.
These charges have garnered much attention recently, because they, and 
related quantities on black-hole horizons, were proposed as a type of 
``soft hair'' on black holes that could be a part of the resolution to the
black-hole-information paradox~\cite{Hawking:2016sgy}. 

The super angular momentum and the supermomentum are also of interest 
because of their relation to gravitational-wave (GW) memory effects.
The first GW memory effect discovered---which in this paper we will simply 
refer to as \textit{the} GW memory effect\footnote{There seem to be two 
competing naming systems for GW memory effects: one is based on the type of
physical effect that could be measured as a consequence of the GW memory; 
the other employs the name of the flux of the ``conserved'' quantity which 
can act as a source of the corresponding memory effect.
Thus, the two nomenclatures would suggest calling it the displacement memory
(as in~\cite{Nichols:2017rqr}) or the 4-momentum (or supermomentum) memory 
(a name that, as far as we can tell, has never been used).
However, because this is the first memory effect discovered, we will opt 
against adding cumbersome modifiers and simply refer to it as \textit{the} GW 
memory effect (and we will typically drop the emphasis on the word ``the'' 
hereafter).}---is characterized by a nonzero change in the GW strain 
between early and late times. 
In an idealized detector composed of freely falling test masses, the GW memory
causes the proper distance of the masses before and after the GWs have passed
through the detector to differ.
The GW memory was initially computed within the context of linearized 
gravity by Zel'dovich and Polnarev~\cite{Zeldovich1974}, and it was 
subsequently computed in full (nonlinear) general relativity by
Christodoulou~\cite{Christodoulou1991}.
Note, however, that the idea of a nonlinear GW memory effect dates back 
(at least) to Payne~\cite{Payne:1984ec} (including the notion that the memory 
is related to supertranslation symmetries and to Weinberg's soft 
theorem~\cite{Weinberg:1965nx})
as well as to an unpublished habilitation thesis, of which certain results were
later published in~\cite{Blanchet1992} (see~\cite{Blanchet:1996yd} for more
detail).\footnote{It seems plausible to argue that the GW memory in full 
general relativity was previously realized as a possibility by Newman and 
Penrose (see the discussion in~\cite{Newman:1966ub}); however, we will not 
attempt to settle the question of the first reference to the GW memory effect
here.}
The sources of the GW memory are changes in the supermomentum 
charges and in the quadrupole and higher-multipole moments of the flux 
of 4-momentum radiated in massless fields and GWs (see,
e.g.,~\cite{Bieri:2013ada,Strominger:2014pwa,Flanagan:2015pxa}).

Pasterski \textit{et al.}~\cite{Pasterski:2015tva} also realized that there
can be a new kind of GW memory effect, which they called the 
\textit{spin memory effect}.
The spin memory is characterized by a change in the time integral of the
magnetic-parity part of the GW strain.\footnote{By ``magnetic-parity part,''
we mean the part that can be decomposed into magnetic-parity tensor 
spherical harmonics (see, e.g.,~\cite{Thorne1980} for a review of these
harmonics).
This turns out to be equivalent to the part parameterized by the scalar 
function $\Psi$ in Eq.~\eqref{eq:CABdecomp}.
It is also sometimes called just the ``magnetic part,'' for short.}
It can be measured by a Sagnac detector following a particular accelerating 
trajectory~\cite{Pasterski:2015tva} or by a family of freely falling observers 
surrounding a source of GWs~\cite{FlanaganInPrep}.
The sources of the spin memory are changes in the superspin charges or in the
quadrupole and higher-multipole moments of the flux of intrinsic part of the
angular momentum carried by massless fields and GWs~\cite{Flanagan:2015pxa}.
The spin memory also has a signature in the gravitational waveform from 
compact binaries that could be detected by third-generation GW 
observatories~\cite{Nichols:2017rqr},
such as the Einstein Telescope~\cite{Punturo:2010zz} and Cosmic 
Explorer~\cite{Evans:2016mbw}.

There has not yet been any discussion of a memory effect related to changes in 
the quadrupole and higher multipole moments of the flux of the CM portion of 
the angular momentum or in the super-CM charges.
We find that there can be such an effect, which we call the 
\textit{center-of-mass (CM) memory effect}.\footnote{Note that we follow
the convention of naming the memory effect after the type of charge that can
generate the effect when it varies in time.
The primary reason for this is to maintain a parallel with the naming of the
spin memory effect.
A secondary reason is that the measurable effect related to the CM memory is 
somewhat involved (as we discuss later), and it does not lend itself to a 
simple name.}
Defining this effect, understanding its properties, and computing the
effect from nonspinning, quasicircular compact binaries are all goals of this 
paper.
To help reach these goals, we will also need to discuss the properties of the 
flux of CM angular momentum and the context in which the CM memory effect is 
defined.
We organize the discussion of these topics as follows.

In Sec.~\ref{sec:CMflux}, we review some properties of the flux of (super)
angular momentum in asymptotically flat spacetimes.
We first provide some background on the Bondi-Sachs framework, the space 
of stationary and nonradiative solutions of Einstein's equations in 
asymptotically flat spacetimes, and BMS symmetries and their corresponding 
charges and fluxes.
We then discuss how changes in the (super) angular momentum transform under 
supertranslations and how they can be interpreted physically.
Even for spacetimes that start in a stationary state, radiate, and then settle 
to a different stationary state (a stationary-to-stationary transition), the 
changes in the charges can transform nontrivially.
We also give an expression for the flux of CM angular momentum carried by GWs,
when the GW strain is expanded in a set of radiative multipole moments.

In Sec.~\ref{sec:CMmemory}, we introduce the CM memory effect,
we discuss the context in which it is defined, and we show that it is 
invariant under infinitesimal BMS supertranslation symmetry transformations.
We also give an expression for the CM memory effect in terms of multipole
moments of the GW strain in this part.

The results of Secs.~\ref{sec:CMflux} and~\ref{sec:CMmemory} are then used in 
Sec.~\ref{sec:CMpn} to compute the leading-order expressions for the CM memory 
effect and flux of CM angular momentum for nonspinning, quasicircular compact 
binaries in the post-Newtonian (PN) approximation.
We find that both equal- and unequal-mass binaries have a CM memory effect,
but the leading-PN-order sources of these memory effects come from the ordinary
and null parts of the memory, respectively (using the terminology of Bieri and 
Garfinkle~\cite{Bieri:2013ada}).
When we estimate the amplitude of the part of the gravitational waveform 
responsible for the CM memory, we find that both the null and the ordinary 
parts will be unlikely to be observed (though for different reasons), even 
for the next generation of ground-based GW detectors, such as the Einstein 
Telescope or Cosmic Explorer.
We conclude in Sec.~\ref{sec:Conclusions}.

Throughout this paper we use units in which $G=c=1$, and we use the 
conventions for spacetime indices and metric and curvature tensors given
in~\cite{Wald:1984rg}.

\section{\label{sec:CMflux} Properties of the flux of (super) angular momentum}

Before discussing the properties of the flux of (super) angular momentum and 
the interpretation of the CM part of the flux, we briefly review a few features
of the Bondi-Sachs framework that will be needed throughout this paper.

\subsection{Aspects of the Bondi-Sachs framework}

The metric of asymptotically flat spacetimes can be expressed in Bondi 
coordinates, $(u,r,\theta^A)$ (where $A=1,2$).
These coordinates are a retarded time ($u$), an affine parameter along 
outgoing null rays as well as an areal radius ($r$), and coordinates 
on a 2-sphere ($\theta^A$).
The general form of the metric and the corresponding Einstein equations were
derived assuming axisymmetry in~\cite{Bondi1962}.
Subsequently, in Ref.~\cite{Sachs1962a}, Einstein's equations without imposing
axisymmetry were given in part; the full expressions for the hypersurface and
evolution equations in vacuum (and with respect to a particular 
parameterization of Bondi-Sachs coordinates) were given 
in~\cite{VanDerBurg1966}.
The hypersurface and evolution equations with matter sources (and in a 
covariant notation with respect to the 2-sphere cross sections) were written
later in~\cite{Winicour1983}.
We will not give these (somewhat lengthy) expressions here; however, 
we will briefly discuss the structure of Einstein's equations as
elaborated in these references.

Of the ten components of Einstein's equations, four take the form of 
``hypersurface'' equations, which do not involve $u$ derivatives, and which
constrain different metric functions on hypersurfaces of constant $u$.
Two other components are evolution-type equations for the 
transverse-traceless parts of the metric.
The final four components are sometimes called the ``conservation'' equations,
though one component is trivially satisfied.
The remaining three components have the property that if they are satisfied
at a fixed value of $r$ on an outgoing null cone in a Bondi coordinate 
chart, then they are satisfied for all such values of $r$.
This follows from the contracted Bianchi identities (which are equivalent to 
local stress-energy conservation for spacetimes with matter sources).

We next briefly review the procedure involved in the derivation of the 
components of Einstein's equations that we will need in the discussion below.
We start from the Bondi-Sachs metric, which we write as
\begin{align}
ds^2  = & - U e^{2 \beta} du^2 - 2 e^{2 \beta} du dr  \nonumber \\
& + r^2 \gamma_{AB} (d\theta^A - U^A du) (d\theta^B - U^B du) \, .
\end{align}
We then assume that the functions $U$, $\beta$, $U^A$ and $\gamma_{AB}$ 
can be expanded in a series in $1/r$ with the asymptotic fall-off 
conditions given in~\cite{Bondi1962}.
When the spacetime contains matter sources, it is also necessary to assume 
fall-off conditions on the stress-energy tensor $T_{ab}$.
We use those discussed in~\cite{Flanagan:2015pxa}, which are based
on the stress-energy tensor of a radiating scalar field in flat spacetime:
\begin{subequations}
\begin{align}
T_{uu} & = r^{-2} \hat T_{uu}(u,\theta^A) + O(r^{-3}) \, , \\
T_{uA} & = r^{-2} \hat T_{uA}(u,\theta^B) + O(r^{-3}) \, , \\
T_{rA} & = r^{-3} \hat T_{rA}(u,\theta^B) + O(r^{-4}) \, , \\
T_{rr} & = r^{-4} \hat T_{rr}(u,\theta^A) + O(r^{-5}) \, , \\
(T_{AB})^{\mathrm{TF}} & = r^{-2} \hat T_{AB}(u,\theta^C) + O(r^{-3}) \, .
\end{align}
\label{eq:TabVals}%
\end{subequations}
The superscript ``TF'' means to take the trace-free part of the expression
on the left-hand side of the equation with respect to the metric on the 
2-sphere, $h_{AB}$.
Note that local stress-energy conservation requires that the functions
$\hat T_{rr}$ and $\hat T_{rA}$ be related by
\begin{equation}
\hat T_{rA}(u,\theta^B) = \check T_{rA}(\theta^B) 
- \frac 12 D_A T_{rr}(u,\theta^B)
\end{equation}
(see, e.g.,~\cite{Flanagan:2015pxa}).
The derivative operator $D_A$ is the Levi-Civita connection compatible with
the metric $h_{AB}$.

The hypersurface-type components of Einstein's equations can then be applied
to determine the precise form of the expansion of the functions 
$U$, $\beta$, $U^A$ and $\gamma_{AB}$ in a series in $1/r$.
At the accuracy in $1/r$ needed for the discussion of Einstein's equations 
below, these functions are given by
\begin{subequations}
\begin{align}
& \! \! \gamma_{AB} = h_{AB} \bigg( 1 + \frac{1}{4r^2} C_{CD}C^{CD} 
+ \frac{1}{2r^3} \mathcal D_{CD} C^{CD} \bigg) \nonumber \\
& \! \! \phantom{\gamma_{AB} =} + \frac 1 r C_{AB} 
+ \frac 1{r^2} \mathcal D_{AB} + \frac 1{r^3} \mathcal E_{AB} + O(r^{-4}) 
\, , \\
& \! \! U^A = -\frac{1}{2r^2} D_B C^{AB} + \frac 1{r^3} \bigg[-\frac 23 N^A
+ \frac 1{16} D^A(C_{BC} C^{BC})  \nonumber \\
& \! \! \phantom{U^A=} + \frac 12 C^{AB} D^C C_{BC} \bigg] + O(r^{-4}) \, , \\
& U  = 1 - \frac{2 m}r + O(r^{-2}) \, ,\\
& \beta = -\frac 1{r^2} \left(\pi \hat T_{rr} + \frac 1{32} C_{AB} C^{AB}
\right) + O(r^{-3}) \, .
\end{align}
\label{eq:BondiPotentials}%
\end{subequations}
In the expressions above, all the scalars and tensors on the right-hand side 
are functions of the coordinates $(u,\theta^A)$, which have been omitted to 
make the notation more compact; also all capital Latin indices are raised and 
lowered with the metric $h^{AB}$ and its inverse.
The tensors $C_{AB}$, $\mathcal D_{AB}$, and $\mathcal E_{AB}$ in the 
expansion of $\gamma_{AB}$ are symmetric and trace free.
This, as well as the form of the term proportional to $h_{AB}$, is required to 
satisfy the determinant condition of Bondi gauge:
$\partial_r \det(\gamma_{AB}) = 0$.
Two of the hypersurface-type components of Einstein's equations also require 
that $D^B \mathcal D_{BA} = -8\pi \check T_{rA}$.
The two additional functions $m(u,\theta^A)$ and $N_A(u,\theta^B)$ are often
called the Bondi mass and angular-momentum aspects, respectively.
We use a convention for the angular-momentum aspect like that used by 
Sachs~\cite{Sachs1962a}, in which it is proportional to the $1/r^4$ parts of 
certain components of the Riemann tensor.

The three nontrivial conservation components of Einstein's equations 
require that the Bondi mass and angular-momentum aspects satisfy the 
following equations:
\begin{subequations}
\begin{align}
\dot m = & -4\pi \hat T_{uu} - \frac 18 N_{AB} N^{AB} + \frac 14 D_A D_B N^{AB}
\, ,\\
\dot N_A = & - 8 \pi \hat T_{uA} + \pi D_A \partial_u \hat T_{rr} + D_A m 
\nonumber \\
& + \frac 14 D_B D_A D_C C^{BC} - \frac 14 D_B D^B D^C C_{CA} \nonumber \\
& + \frac 14 D_B (N^{BC} C_{CA}) + \frac 12 D_B N^{BC} C_{CA} \, .
\end{align}
\label{eq:Aspects}%
\end{subequations}
The dot over the variables on the left-hand side is a short-hand notation
for $\partial_u$.
We define the news tensor as $N_{AB} = \partial_u C_{AB}$ (twice that defined
in~\cite{Bondi1962}).
The news tensor is a quantity that arises from solving the evolution equations 
for the traverse-traceless components of Einstein's equations at leading order 
in $1/r$.
The news tensor is unconstrained, but it can be shown that it vanishes when 
the spacetime is not radiating GWs~\cite{Geroch1977}.

Expanding the evolution-type components of Einstein's equations at higher order
in $1/r$, leads first to the equation $\dot{\mathcal D}_{AB} = 0$, which is 
consistent with the hypersurface-type equations 
$D^B \mathcal D_{BA} = -8\pi \check T_{rA}$.
When the spacetime is vacuum, it follows that $\mathcal D_{AB} = 0$.
The tensor $\mathcal E_{AB}$ satisfies a nontrivial evolution equation:
\begin{align}
\dot{\mathcal E}_{AB} = & -4\pi \hat T_{AB} - 2\pi(\partial_u \hat T_{rr}) 
C_{AB} -\frac 12 \mathcal D_{AB} + \frac 12 m C_{AB}  \nonumber \\
& + \pi \left(D_A D_B - \frac 12 h_{AB} D^2\right) \hat T_{rr}
+ \frac 13 D_{(A} N_{B)} \nonumber \\
& - \frac 16 h_{AB} (D_C N^C) + \frac 14 C_{AB} (N_{CD} C^{CD}) \nonumber \\
& - \frac 18 {\epsilon_A}^C C_{CB} (\epsilon_{DE} D^E D_C C^{CD}) \, .
\label{eq:EABevolve}
\end{align}
A closely related equation in axisymmetry and in vacuum appears in the 
paper~\cite{Bondi1962}. 
Restricting Eq.~\eqref{eq:EABevolve} to vacuum, it is equivalent to an 
equation derived by van der Burg~\cite{VanDerBurg1966} after taking into 
account differences in notation and convention used (Sachs~\cite{Sachs1962a}
also derives a related equation, but does not present all the nonlinear
terms).
The linearized limit of Eq.~\eqref{eq:EABevolve} also agrees with the 
nonvacuum expression given in, e.g.,~\cite{Madler:2016ggp}.
Because the tensor $\mathcal E_{AB}$ is closely related to the Newman-Penrose 
scalar $\psi_0$~\cite{Newman:1961qr} (discussed in~\cite{Sachs1962a}), it is
also closely related to evolution equations for this scalar (see, for example,
the review~\cite{Adamo:2009vu}).

\subsection{Stationary and nonradiative regions and transitions between these
regions}

For computing memory effects, we specialize to asymptotically flat spacetimes 
that begin in a stationary or a nonradiative ($N_{AB}=0$) state, radiate GWs 
and massless fields, and then settle into a different nonradiative or 
stationary state.
We will often make the further assumptions that the initial stationary or 
nonradiative region is in vacuum ($T_{ab} = 0$), the radiative region of the 
spacetime is not in vacuum [and the stress-energy tensor satisfies the 
conditions in Eq.~\eqref{eq:TabVals}], and the final stationary or 
nonradiative region also is in vacuum. 
Einstein's equations in~\eqref{eq:Aspects} and~\eqref{eq:EABevolve} constrain
the form of the Bondi-metric functions $m$, $N_A$, and 
$\mathcal E_{AB}$ in stationary or nonradiative regions, which restricts the
space of solutions to Einstein's equations therein.
However, it does not imply that a given set of astrophysical sources will
necessarily realize the full space of solutions consistent with the vacuum and
stationary or nonradiative conditions.

This type of issue (as it relates to the GW memory effect) was discussed by 
Frauendiener~\cite{Frauendiener1992}.
From the perspective of Einstein's equations, the news tensor can be an 
arbitrary function $N_{AB}(u,\theta^C)$, and in a nonradiative-to-nonradiative
transition, the memory can have any amplitude and angular dependence (this 
should hold for the changes in $m$, $N_A$, and $\mathcal E_{AB}$, too).
However, from the perspective of solving a specific initial-value problem
for a certain astrophysical source, the news tensor cannot be specified freely; 
rather, it follows from the dynamics of the source.
The Bondi news tensor can be determined through some sort of matching procedure 
analytically (e.g., through post-Newtonian-expanded, multipolar-post-Minkowski 
calculations~\cite{Blanchet:2013haa}) or numerically (e.g., through 
Cauchy-characteristic extraction~\cite{Bishop:1996gt}).
For the specific systems treated in this paper (inspiraling compact binaries, 
and particularly binary-black-hole mergers), the allowed values of the memory, 
and the changes in $m$, $N_A$, and $\mathcal E_{AB}$ are more restricted than 
those allowed by the general solutions of Einstein's equations in a 
stationary or nonradiative region.\footnote{This issue can be recast in terms 
of how the conservation-type components of Einstein's equations are treated.
These equations are automatically satisfied for all $r$ on an outgoing null 
cone in a Bondi coordinate patch, so long as they are satisfied on some 
2-sphere of fixed $r$.
One choice for this 2-sphere is at infinite radius (i.e., at future
null infinity).
At this boundary of an asymptotically flat spacetime, it is possible to
allow for any value of the news tensor $N_{AB}$, because quantities at null 
infinity can be defined without reference to the interior of the spacetime.
From this perspective, however, it is not clear if these values of the Bondi 
news tensor correspond to any astrophysical solution of Einstein's equations 
in the interior of the spacetime.
The other viewpoint, which fits more with the aims of this paper, is to allow
the Bondi functions to satisfy the conservation-type components of Einstein's 
equations at finite $r$ and to determine their evolution by matching to a 
specific initial-value (Cauchy) solution for a given system (as described, 
e.g., in~\cite{Bishop:1997ik}).}
In this paper, we will focus on this latter perspective, because we are 
ultimately interested in GW memory effects arising from the inspiral and 
merger of nonspinning compact binaries.
Nevertheless, we will first describe the general solutions of Einstein's 
equations in nonradiative and stationary regions, to make clear the
types of restrictions we are making in specializing to particular sources.

In a nonradiative and vacuum region, the first line of Eq.~\eqref{eq:Aspects} 
requires that the Bondi mass aspect is independent of $u$, so that it is just 
a function of angular coordinates, $m(\theta^A)$.
From the other lines of Eq.~\eqref{eq:Aspects}, it then follows that 
$N_A$ can have a piece that depends linearly on $u$.
The electric-parity part of $N_A$ depends on $D_A m$, while the magnetic-parity
part depends on the magnetic-parity part of $C_{AB}$.
Although the magnetic-parity part vanishes in stationary regions 
(see~\cite{VanDerBurg1966,Newman:1968uj}), it need not vanish in nonradiative 
regions.
$N_A$ can also have a part that is independent of $u$ (with both electric and 
magnetic parities).
Finally, from Eq.~\eqref{eq:EABevolve}, it then implies that $\mathcal E_{AB}$
can have terms proportional to $u^2$, $u$, and independent of $u$ 
with both electric and magnetic parities, in a nonradiative, vacuum region.
Summarizing these results by explicitly solving Eqs.~\eqref{eq:Aspects} 
and~\eqref{eq:EABevolve} in such a region, we find that
\begin{subequations}
\begin{align}
m = & m(\theta^A) \, ,\\
N_A = & u D_A m + \frac u 4 (D_B D_A D_C C^{BC} - D^2 D^B C_{AB} ) \nonumber \\
& + N_A^{(0)}(\theta^B) \, ,\\
\mathcal E_{AB} = & \frac{u^2}{24} [4 D_A D_B m - 2 D^2 m h_{AB} + 
D_B D_A D_C C^{BC} \nonumber \\
& - D^2 D^B C_{AB}] + \frac{u}{2} m C_{AB} 
+ \frac{u}{6} (2 D_{(A} N_{B)}^{(0)} \nonumber \\
& - D^C N_C^{(0)} h_{AB} ) 
- \frac{u}{8}  {\epsilon_A}^C C_{CB} (\epsilon_{DE} D^E D_C C^{CD}) \nonumber \\
& + \mathcal E_{AB}^{(0)} (\theta^C) \, .
\end{align}
\label{eq:AspectsNonRad}%
\end{subequations}
Recall that while Eq.~\eqref{eq:AspectsNonRad} is the most general solution 
for $m$, $N_A$, and $\mathcal E_{AB}$ consistent with a nonradiative and vacuum
region of future null infinity, it is not clear if the nonradiative regions of 
a specific astrophysical system, such as a merging compact binary, will realize
this level of generality.

Stationary vacuum regions, for example, have frames in which the Bondi metric 
functions are independent of $u$~\cite{VanDerBurg1966,Newman:1968uj}.
Applying this condition to Eq.~\eqref{eq:AspectsNonRad}, we find that
$m$ is a constant, the magnetic-parity part of $C_{AB}$ is zero, and 
$N_A^{(0)}$ is composed of both $l=1$ vector spherical harmonics and $l>1$
harmonics that satisfy
$2D_{(A}N_{B)}^{(0)} - (D^C N_C^{(0)}) h_{AB} = -3 m C_{AB}$.
These frames can then be transformed to the ``canonical'' frame described 
in~\cite{Flanagan:2015pxa}, in which $m$ is constant, $C_{AB}=0$, and 
$N_A^{(0)}$ is composed of $l=1$ magnetic-parity vector harmonics.

Because our primary focus in this paper is on merging compact binaries composed
of black holes, we will need to know the properties of the nonradiative regions 
for these binaries at early and late times in their evolution.
At early times, the binaries can be approximated well by PN theory.
One assumption in this approximation is that there is a (finite) time before
which the system was stationary in the past 
(see, e.g.,~\cite{Blanchet:2013haa}).
This could correspond to a time early in the evolution of the binary, when the 
binary's components are sufficiently widely separated and slowly moving 
that the system can be treated as stationary.
The outcome of a binary-black-hole merger is a stationary black hole.
For studying binary-black-hole mergers, therefore, it should be 
sufficient to consider stationary-to-stationary transitions.
It is also important to briefly describe the types of restrictions assuming a
stationary-to-stationary transition will cause, so as to better understand
the generality of our results.

For simplicity, in most of the subsequent calculations and discussion, we will 
assume that the initial stationary frame is the canonical frame of the system.
At late times, the stationary frame will generally not be the canonical frame,
but one that differs from the canonical frame by a BMS transformation (which
can be decomposed into a rotation, followed by a boost, and then a 
supertranslation).
From these properties of the initial and final frames, we anticipate that 
there will be two different types of restrictions from assuming a 
stationary-to-stationary transition.

The first is that the magnetic-parity part of the shear will vanish in both 
stationary regions (although, in general $m$ will not be constant and $N_A$ 
will not consist of just $l=1$ magnetic-parity vector harmonics in the final 
stationary region).
This does not seem to be a very strong restriction, because M\"adler and
Winicour~\cite{Madler:2016ggp} have shown that there is no magnetic-parity
memory effect in the absence of incoming radiation or time-dependent, 
anisotropic, magnetic-parity material stresses near null infinity 
(which compact binaries are generally not expected to have, for example).
Several common classes of stress-energy tensors also do not give rise to 
magnetic-parity memory~\cite{Madler:2016ggp}.
The second type of restriction relates to the ordinary part of the GW memory
(using the terminology of~\cite{Bieri:2013ada}).
Assuming a stationary-to-stationary transition makes the ordinary part of
the GW memory a function of just the change in the 4-momentum radiated by 
the spacetime.
It, therefore, would exclude certain physically relevant systems, like 
the gravitational scattering of astrophysical objects considered 
in~\cite{Zeldovich1974}.
Note, however, that the assumption of a stationary-to-stationary transition
does not have a significant effect on the null part of the GW memory (neither 
the linear nor the nonlinear parts).

Finally, because the set of stationary-to-stationary transitions is contained
within the larger set of nonradiative-to-nonradiative transitions, imposing 
the former assumption will generally restrict the types of possible memory
effects.
Because stationary-to-stationary transitions contain an interesting set of 
physical systems (compact-binary mergers), it has sufficient generality to
allow for some nontrivial memory effects (even if they are not the most 
general effects possible).
Having elaborated our assumptions and their consequences, we next discuss BMS 
symmetries and their conjugate charges and fluxes.

\subsection{Symmetries, charges, and fluxes}

The vector fields at future null infinity that define the (extended) BMS 
algebra, $\vec \zeta$, are parameterized by a scalar function 
$\alpha(\theta^A)$ and a vector on the 2-sphere $Y^A(\theta^B)$ as follows:
\begin{equation}
\vec \zeta = [\alpha(\theta^A) + u D_A Y^A(\theta^B)/2]\vec \partial_u + 
Y^A(\theta^B) \vec \partial_A \, .
\label{eq:BMSsymmetry}
\end{equation}
The quantity $\alpha(\theta^A)$ is a smooth function that corresponds to a
supertranslation, and $Y^A(\theta^B)$ are $l=1$ vector spherical harmonics,
for the standard BMS group.
For the extended BMS algebra~\cite{Barnich:2011mi}, $Y^A$ are elements of a 
Virasoro algebra, or for the generalized BMS group~\cite{Campiglia:2014yka},
they are smooth vector fields on the 2-sphere.
The standard and extended BMS symmetries at null infinity can be defined at 
finite $r$ in Bondi coordinates by requiring that the spacetime metric
continues to satisfy the Bondi gauge conditions and the same scaling with $r$ 
under pullback along the symmetry vector fields.
The vector fields in~\eqref{eq:BMSsymmetry} have a series expansion in $1/r$ 
in the interior of the spacetime, and the Bondi metric functions ($C_{AB}$, 
$m$, and $N_A$) transform nontrivially under these (extended) BMS symmetries.
The formulas for the BMS vector fields and the transformations of the 
Bondi functions are given, for example, in~\cite{Chrusciel:2002}.

For most of the computations in this paper, we are interested in how the
Bondi functions transform under supertranslations in vacuum and in stationary 
or nonradiative regions of the types described in the previous subsection.
Specializing the results in~\cite{Flanagan:2015pxa}, for example, we find that
$m$ is invariant under supertranslations and
\begin{subequations}
\begin{align}
\delta C_{AB} = & (-2 D_A D_B + h_{AB} D^2) \alpha \equiv -2 C_{AB}^{(\alpha)}
\\
\delta N_A = & \alpha D_A m + 3mD_A \alpha + \frac 14 C_{AB} D^B D^2 \alpha
\nonumber \\
& -\frac 34 D_B\alpha(D^B D^C C_{CA} - D_A D_C C^{BC}) \nonumber \\
& + \frac 38 D_A (C^{BC} C_{BC}^{(\alpha)} ) 
+ \frac 12 C_{AB}^{(\alpha)} D_C C^{BC} 
\end{align}
\label{eq:TransformNA}%
\end{subequations}
(the first line can be found from the results in~\cite{Sachs1962a} 
or~\cite{Newman:1966ub}).
In the equation above, we have introduced the notation $C_{AB}^{(\alpha)}$
to denote the electric-parity part of the shear generated by a scalar 
``potential'' $\alpha(\theta^A)$.

The (super) angular momentum in a vacuum, nonradiative region of null infinity, 
on a cut $\mathcal C$ of constant $u=u_0$, is given by
\begin{align}
Q[\vec\zeta_Y;\mathcal C] & = \frac 1{128\pi} \int d^2\Omega 
Y^A[16(N_A - u_0 D_A m) \nonumber \\
& -D_A(C_{BC} C^{BC}) - 4 C_{AB} D_C C^{BC}] \, ,
\label{eq:ChargeQy}
\end{align}
where by $\vec\zeta_Y$, we mean a BMS vector field with $\alpha=0$, and which
is thus parameterized by the vector on a 2-sphere, $Y^A$.
The prescription to compute this charge corresponding to a vector field 
$Y^A(\theta^B)$ is outlined in~\cite{Flanagan:2015pxa}, which is based on the
procedure in~\cite{Wald:1999wa} (and which gives equivalent results to those
defined via a different procedure in~\cite{Barnich:2011mi}, in the nonradiative
and vacuum regions treated here).

The integral of the flux of (super) angular momentum between two cuts 
$\mathcal C_1$ and $\mathcal C_2$ in vacuum, nonradiative regions given by 
$u=u_1$ and $u=u_2$, respectively, is 
\begin{align}
\Delta \tilde Q[\vec\zeta_Y;\mathcal C_2, \mathcal C_1] & = -\frac 1{64\pi} 
\int_{u_1}^{u_2} \! du \int d^2\Omega [ u D_A(2 D_B D_C N^{BC} 
\nonumber \\
& - N_{BC} N^{BC} -32\pi \hat T_{uu}) + D_A (C_{BC}N^{BC}) \nonumber \\
& + 2 N^{BC} D_A C_{BC} -4 D_B(N^{BC}C_{AC}) \nonumber \\
& + 64\pi \hat T_{uA}] Y^A \, .
\label{eq:FluxNoSM}
\end{align}
It was shown in~\cite{Flanagan:2015pxa} that the changes in the charges between
the two cuts $\mathcal C_1$ and $\mathcal C_2$ do not equal the integral of 
the flux in Eq.~\eqref{eq:FluxNoSM} for the meromorphic super-rotation vector 
fields $Y^A$ (i.e., when $Y^A$ is not one of the six generators of the
Lorentz group).
To restore equality for these extended BMS symmetries, an additional term of
the form
\begin{align}
\Delta \mathcal F[\vec\zeta_Y;\mathcal C_2, \mathcal C_1] \equiv & 
\frac 1{32\pi} \int_{u_1}^{u_2} du \int d^2\Omega Y^A \epsilon_{AB} 
\epsilon^{CD} \nonumber \\
& \qquad \times D^B D_D D^E C_{CE} \, .
\label{eq:FluxSMterm}
\end{align}
must be added.
The change in the charges is then given by
\begin{equation}
Q[\vec \zeta_Y;\mathcal C_2] - Q[\vec \zeta_Y;\mathcal C_1] = 
\Delta \tilde Q[\vec \zeta_Y;\mathcal C_2, \mathcal C_1] 
- \Delta \mathcal F[\vec \zeta_Y;\mathcal C_2, \mathcal C_1] \, .
\end{equation}
We reiterate that the term 
$\Delta \mathcal F[\vec \zeta_Y;\mathcal C_2, \mathcal C_1]$
vanishes for the standard BMS group; it is only needed for the additional 
elements of the extended BMS algebra.
The term $\Delta \mathcal F[\vec \zeta_Y;\mathcal C_2, \mathcal C_1]$ is also
closely related to the spin memory effect of~\cite{Pasterski:2015tva},
as discussed in~\cite{Flanagan:2015pxa}.

It will be convenient to define a quantity that is equal to the change in the 
charges:
\begin{equation}
\Delta Q[\vec \zeta_Y;\mathcal C_2, \mathcal C_1] \equiv 
Q[\vec \zeta_Y;\mathcal C_2] - Q[\vec \zeta_Y;\mathcal C_1]
\label{eq:ChargeFlux}
\end{equation}
After some algebra (described in~\cite{Flanagan:2015pxa}), it was shown that 
the change in the charges can be written as
\begin{align}
\! \Delta Q[\vec \zeta_Y;\mathcal C_2, \mathcal C_1] & = -\frac 1{64\pi} \!
\int_{u_1}^{u_2} \! du \int \! d^2\Omega [u D_A ( 2D_B D_C N^{BC} 
\nonumber \\ 
& - N_{BC} N^{BC} - 32\pi \hat T_{uu}) + C^{BC} D_B N_{AC} \nonumber \\
& - N^{BC} D_B C_{AC} + 3(N_{AB} D_C C^{BC} \nonumber \\
& - C_{AB} D_C C^{BC})  +64\pi\hat T_{uA} + 16\pi \partial_u \hat T_{rA}
\nonumber \\
& + 2 \epsilon_{AB} \epsilon^{CD} D^B D_D D^E C_{CE} ] Y^A \, .
\label{eq:FluxFull}
\end{align}

\subsection{Transformation properties of (super) angular momentum under
supertranslations}

We now point out a few features of the (super) angular momentum charges and
fluxes that we have not seen discussed explicitly elsewhere, but which may
be related to two other aspects of the charges and fluxes that have been 
previously noted.
The first is that nonlinear terms involving the shear in the super angular 
momentum can make it behave nontrivially: for example, it can be nonvanishing 
in spacetimes that are flat aside from a defect at the 
origin~\cite{Compere:2016jwb}.
The second is the observation that the flux of angular momentum will 
depend upon nonradiative (or ``Coulombic'') parts of the Bondi metric
functions and stress-energy tensor~\cite{Ashtekar:2017wgq}.

To illustrate the transformation properties of the (super) angular momentum, 
we will examine the same stationary-to-stationary transition from the 
perspective of two different Bondi frames.
For the first frame, we use the canonical frame associated with the initial
stationary region.
Constructing this frame fixes all the degrees of freedom in the BMS group 
except for a global SO(3) rotation and a time translation
(a BMS transformation with $\vec \zeta = u_0 \vec \partial_u$, 
for a constant, $u_0$).
We denote by $\mathcal C_1$ a cut corresponding to a retarded time $u=u_1$ in 
the initial stationary region and by $\mathcal C_2$ a cut of constant $u=u_2$
in the latter stationary region.
For the second Bondi frame, we will consider one that is supertranslated from 
the canonical Bondi frame of the initial stationary region by an amount 
$\alpha$.
Because $u'=u+\alpha$, we will denote the cuts by $\mathcal C_1'$ and 
$\mathcal C_2'$, which correspond to 2-sphere cross sections of constant 
$u'=u_1'$ and $u'=u_2'$, respectively.
Finally, we also assume that the spacetime has GW memory, which is determined
by a potential $\Delta\Phi(\theta^A)$ and which is given by
\begin{align}
\Delta C_{AB} & = C_{AB}(u_2) - C_{AB}(u_1) \nonumber \\
& = \frac 12 (2D_A D_B - h_{AB} D^2) \Delta\Phi \, .
\label{eq:MemDeltaPhi}
\end{align}
The memory is invariant under supertranslations (i.e., is equivalent to the 
related quantity measured at the times $u_2'$ and $u_1'$).
We will treat the supertranslation, $\alpha$, as small, and we will compute
the transformation of the charges to linear order in $\alpha$.
We will not linearize with respect to the potential $\Delta \Phi$ that 
determines the GW memory.

We are particularly interested in comparing the changes in the charges between 
the cuts $\mathcal C_1$ and $\mathcal C_2$ with those between the cuts 
$\mathcal C_1'$ and $\mathcal C_2'$.
Performing such a comparison is somewhat subtle, because the 
(extended) BMS vector fields corresponding to (super) Lorentz transformations 
on cuts of constant $u$ and $u'$ are different.
Namely, the quantity
\begin{equation}
\vec \zeta_Y = \frac 12 u D_A Y^A \vec\partial_u + Y^A \vec \partial_A 
\label{eq:ZetaYvec}
\end{equation}
and the equivalent vector fields adapted to the cuts of constant $u'$ differ 
by a supertranslation (e.g.,~\cite{Sachs1962b} and~\cite{Barnich:2011mi}).
The charges associated with these two vector fields will therefore include 
different amounts of supermomentum.
While this is to be expected, the difference in the charges arising from the 
dependence of the charges on the cut will mix with the difference that comes 
from the dependence of the charges on the generators adapted to those cuts.
Instead, we will compute the change in the charges between the cuts of 
constant $u'$ with the generators adapted to cuts of constant $u$.
The vector field $\vec \zeta_Y $ expressed in terms of the primed
coordinates is given by
\begin{equation}
\vec \zeta_Y =  \left[\frac 12 (u' - \alpha) (D_A Y^A) + Y^A D_A \alpha\right]
\vec \partial_{u'} + Y^A \vec \partial_A
\end{equation}
(see, e.g.,~\cite{Madler:2017umy}).

We will now show that in stationary vacuum regions, the (super) angular 
momentum transforms nontrivially under supertranslations (unlike the 
supermomentum, which is supertranslation invariant in this context).
The reason for this is as follows.
Although we use the same BMS vector field, $\vec \zeta_Y$, to compute the 
charges in the cuts defined by $u$ and $u'$, because the cuts
of constant $u'$ are supertranslated from the cuts of constant $u$, the values
of the Bondi metric functions $C_{AB}$ and $N_A$ differ between the two
sets of cuts (even in the vacuum and stationary regions).
In addition, the split of the vector field $\vec\zeta_Y$ into parts tangent
and orthogonal to cuts of $u$ and $u'$ will differ, which will also influence
the value of the charges.
Finally, because the (super) angular momentum charge depends on $C_{AB}$ and 
its derivatives quadratically, it follows from Eqs.~\eqref{eq:TransformNA}
and~\eqref{eq:ChargeQy} that the (super) angular momentum charges that are 
supertranslated from a stationary region in which $C_{AB}=0$ differ from the 
charges that are supertranslated from a frame with a nonzero $C_{AB}$.
While the physical reason for this is not immediately obvious, we 
speculate that these nonlinear terms capture a difference in the ``origin'' 
about which the (super) angular momentum is computed in these two cases.

Let us explicitly compute how this change in the charges produced by a 
supertranslation (which we will denote by 
$\delta Q[\vec\zeta_Y; \mathcal C', \mathcal C]$) will affect the change in 
the (super) angular momentum between the two stationary regions (i.e., 
$\Delta Q[\vec\zeta_Y; \mathcal C_2, \mathcal C_1]$ versus
$\Delta Q[\vec\zeta_Y; \mathcal C_2', \mathcal C_1']$).
In the stationary region including $u_1$ and $u_1'$, because we are working
to linear order in the supertranslation $\alpha$ from the canonical frame,
then the change is similar to a result for the (super) angular 
momentum charges in~\cite{Flanagan:2015pxa}.
To linear order in $\alpha$, we find that
\begin{equation}
\delta Q[\vec\zeta_Y; \mathcal C'_1, \mathcal C_1] = \frac{1}{8\pi} 
\int d^2\Omega (5 Y^A D_A \alpha - \alpha D_A Y^A) m_1 \, ,
\label{eq:deltaQyC1}
\end{equation}
where we have used the notation $m_1 = m(u_1) = m(u_1')$.

Next, let us compute $\delta Q[\vec \zeta_Y; \mathcal C'_2, \mathcal C_2]$.
The expression for this quantity is somewhat lengthier, because we are
allowing $C_{AB}$ to be nonzero at late times (and equal to the GW memory,
$\Delta C_{AB}$, in the cut $u=u_2$).
Using Eqs.~\eqref{eq:TransformNA} and~\eqref{eq:ChargeQy}, we find that to
linear order in $\alpha$
\begin{align}
\delta Q[\vec\zeta_Y; \mathcal C'_2, \mathcal C_2] = & \frac{1}{8\pi} 
\int d^2\Omega (5 Y^A D_A \alpha - \alpha D_A Y^A) m_2  \nonumber \\
& + \frac{1}{64\pi} \int d^2\Omega Y^A [2 \Delta C_{AB} D^B D^2 \alpha - 
\nonumber \\
& 6 D_B \alpha (D^B D^C \Delta C_{CA} - D_A D_C \Delta C^{BC}) \nonumber \\
& + 5 D_A (\Delta C^{BC} C^{(\alpha)}_{BC}) + 8 C_{AB}^{(\alpha)} 
D_C \Delta C^{BC} \nonumber \\
& + 4 \Delta C_{AB} D_C C^{BC}_{(\alpha)} \, .
\label{eq:deltaQyC2}
\end{align}
Given the relationship in Eq.~\eqref{eq:ChargeFlux}, then by construction,
the changes in the charges between the cuts $\mathcal C_1$ and $\mathcal C_2$ 
and the cuts $\mathcal C_1'$ and $\mathcal C_2'$ are related by
\begin{align}
\Delta Q[\vec\zeta_Y; \mathcal C'_2, \mathcal C'_1] = &
\Delta Q[\vec\zeta_Y; \mathcal C_2, \mathcal C_1] + 
\delta Q[\vec\zeta_Y; \mathcal C_2', \mathcal C_2] \nonumber \\
& - \delta Q[\vec\zeta_Y; \mathcal C_1', \mathcal C_1] \, .
\end{align}
Using Eqs.~\eqref{eq:deltaQyC1} and~\eqref{eq:deltaQyC2}, we can compute 
a difference in the changes of the charges,
$\Delta Q[\vec \zeta_Y; \mathcal C'_2, \mathcal C'_1] - 
\Delta Q[\vec \zeta_Y; \mathcal C_2, \mathcal C_1]$, which we find is
\begin{align}
& \Delta Q[\vec \zeta_Y; \mathcal C'_2, \mathcal C'_1] - 
\Delta Q[\vec\zeta_Y; \mathcal C_2, \mathcal C_1] = \nonumber \\
& \frac{1}{8\pi} 
\int d^2\Omega (5 Y^A D_A \alpha - \alpha D_A Y^A) \Delta m  
+ \frac{1}{64\pi} \int d^2\Omega Y^A \times \nonumber \\
& [2 \Delta C_{AB} D^B D^2 \alpha - 
6 D_B \alpha (D^B D^C \Delta C_{CA} - D_A D_C \Delta C^{BC}) \nonumber \\
& + 5 D_A (\Delta C^{BC} C^{(\alpha)}_{BC}) + 8 C_{AB}^{(\alpha)} 
D_C \Delta C^{BC} \nonumber \\
& + 4 \Delta C_{AB} D_C C^{BC}_{(\alpha)} ] \, .
\label{eq:FluxChange}
\end{align}
We defined $\Delta m = m_2 - m_1$ in the expression above.
This result is interesting, because the change in the charges is related to the 
integral of the flux (plus the additional term 
$\Delta \mathcal F[\vec \zeta_Y; \mathcal C'_2, \mathcal C'_1]$).
Thus, while it was not very surprising that the (super) angular momentum 
charges transform under supertranslations, it is more surprising that this
change arising from a BMS transformation does not cancel between early and 
late times in a stationary-to-stationary transition (i.e., the flux 
transforms nontrivially under supertranslations).

From Eq.~\eqref{eq:FluxChange}, it is clear that this lack of cancellation
occurs when the system radiates supermomentum or when there is GW memory.
Thus, the result in Eq.~\eqref{eq:FluxChange} is a combined effect of the GW 
memory, changes in the supermomentum, and the transformation properties of the 
(super) angular momentum under supertranslations.
This is an interesting feature of the change in the (super) angular momentum 
that will be relevant when we discuss the flux of the CM angular momentum in 
the next subsections.
We do not anticipate that it will play an important role for the
CM memory effect: in Sec.~\ref{sec:CMmemory}, we show that the CM memory is
invariant under infinitesimal supertranslations $\alpha$.
It may also be possible to modify this transformation property of the change 
in the charges by an appropriate redefinition of the charges.
Investigating this issue, however, goes beyond the scope of this work.

\subsection{Center-of-mass part of (super) angular momentum and its flux}

In this part, we focus on a few issues that apply specifically to the
(super-) CM part of the angular momentum.
CM angular momentum is the conserved quantity conjugate to Lorentz boost 
symmetries.
In special relativity, it is usually denoted by $K^i$, and it is closely 
related to the mass-weighted CM position, $G^i$.
When there are no external forces, these two quantities satisfy the 
relationships
\begin{equation}
K^i = G^i - t P^i \, , \qquad 
\frac{dG^i}{dt} = P^i \, , \qquad \frac{dK^i}{dt} = 0 
\end{equation}
(see, e.g.,~\cite{deAndrade:2000gf}).
Thus, we see that $K^i$ is the conserved quantity in this context, and
that it represents the mass times the CM position in the center-of-momentum 
frame.
It is also a trivial quantity in this context, because by translating the 
origin of coordinates around which the CM is computed, the CM part of the
angular momentum can be set to zero.

In stationary regions of asymptotically flat spacetimes, the (super-) CM
angular momentum [defined by the integral of the electric-parity part
of the integrand in Eq.~\eqref{eq:ChargeQy} against a vector field $Y^A$]
is again trivial; by performing the BMS transformations needed to reach the
canonical frame, we can make the (super-) CM angular momentum vanish 
(see~\cite{Flanagan:2015pxa}).
We argue below that the change in the (super-) CM angular momentum can be
nontrivial in a stationary-to-stationary transition from the canonical frame
of the first stationary region (in the sense that the CM angular momentum
contains additional information not contained in the changes of other BMS
charges or in the GW memory or spin memory effects, in this context). 
We provide further evidence for this by computing the flux of CM angular
momentum in the PN approximation in Sec.~\ref{sec:CMpn}.
It could be of interest to compare this result to a related calculation of
the flux of CM angular momentum in numerical relativity simulations 
in~\cite{Handmer:2016mls}, though we will not attempt to do this in this paper. 
Instead, we will first point out a few more general features about the CM 
angular momentum and its flux, before we investigate these quantities for 
compact-binary sources in Sec.~\ref{sec:CMpn}.

Because the CM angular momentum, $K^i$, is the mass times the CM 
position in the rest frame of the system (in the context of special relativity,
with no external forces), it is worth briefly discussing the
physical interpretation of this quantity when the CM of the system is changing 
because of radiated linear momentum.
To do so, let us recast Eq.~\eqref{eq:FluxFull} for the change in the charges
in terms of the instantaneous flux on a cut of constant $u$:
\begin{align}
\dot K_{\vec \zeta_Y} = & -\frac 1{64\pi} \int d^2\Omega Y^A 
[u D_A ( 2D_B D_C N^{BC} - N_{BC} N^{BC} \nonumber \\
& - 32\pi \hat T_{uu}) + C^{BC} D_B N_{AC} - N^{BC} D_B C_{AC} \nonumber \\
& + 3(N_{AB} D_C C^{BC} - C_{AB} D_C C^{BC}) \nonumber \\
& +64\pi\hat T_{uA} + 16\pi \partial_u \hat T_{rA}] \, .
\label{eq:FluxBondi}
\end{align}
Note that we have denoted this flux by $\dot K_{\vec \zeta_Y}$ to parallel the
notation commonly used for the CM angular momentum in special relativity.
Integrating the first three terms in Eq.~\eqref{eq:FluxBondi} by parts, we 
find that these terms have exactly the same form as the flux of supermomentum; 
however, instead of a scalar $\alpha(\theta^A)$ appearing in the charge 
integral, it is $u D_A Y^A(\theta^B)/2$.
This is to be expected given that the BMS vector field~\eqref{eq:BMSsymmetry}
contains a sum of both $\alpha$ and $u D_A Y^A /2$ in the $\vec \partial_u$
direction.
The remaining terms in the integrand [which are related to the part 
$\vec Y = Y^A \vec \partial_A$ of the vector field $\vec \zeta_Y$ in 
Eq.~\eqref{eq:ZetaYvec}] have a similar form to the flux of the (super) spin; 
however, they are now the electric-parity part of the integrand, 
rather than the magnetic-parity part.
Because it is the electric-parity part, the term related to the spin memory in 
Eq.~\eqref{eq:FluxSMterm} does not contribute. 
To emphasize the contributions from the two types of terms, we will write the 
instantaneous flux as the sum of two terms as follows:
\begin{equation}
\dot K_{\vec \zeta_Y} = \dot k_{\vec Y} + \frac u2 \dot P_{(D_A Y^A)} \, .
\label{eq:dKduSplit}
\end{equation}
The second term involving $\dot P_{(D_A Y^A)}$ has the same form as the
supermomentum flux (for a scalar function $D_A Y^A$ rather than $\alpha$), and
the quantity $\dot k_{\vec Y}$ contains the remaining terms, which are related 
to the part of $\vec \zeta_Y$ not proportional to $\vec \partial_u$.

Consider now the change in the (super) CM angular momentum in a
stationary-to-stationary transition. 
Given the splitting in Eq.~\eqref{eq:dKduSplit}, this change can be written
as
\begin{equation}
\Delta K_{\vec \zeta_Y}(u_2,u_1) = \Delta k_{\vec Y} +
\int_{u_1}^{u_2} du \frac u2 \dot P_{(D_A Y^A)} \, .
\label{eq:DeltaKsplit}
\end{equation}
Note that this is a specialization of and rewriting of Eq.~\eqref{eq:FluxFull}; 
we have used the notation $\Delta K_{\vec \zeta_Y}(u_2,u_1)$ rather than 
$\Delta Q[\vec\zeta_Y; \mathcal C_2, \mathcal C_1]$ to emphasize that it 
applies specifically to the change of the CM angular momentum.
We also write the vector field as a subscript and use $u_2$ and $u_1$ rather 
than $\mathcal C_2$ and $\mathcal C_1$ to make the notation more compact 
(which will be particularly helpful for when we derive the multipolar 
expansion of the change in the CM angular momentum, which we do in the next 
subsection).
Integrating the second term in Eq.~\eqref{eq:DeltaKsplit} by parts, the change 
has the form
\begin{align}
\Delta K_{\vec \zeta_Y}(u_2,u_1) = & \Delta k_{\vec Y} + \frac 12 \left.
[u P_{(D_A Y^A)}] \right|^{u_2}_{u_1} \nonumber \\
& - \frac 12 \int_{u_1}^{u_2} du P_{(D_A Y^A)} \, .
\label{eq:DeltaKYu1u2}
\end{align}
Thus, we can now better understand the physical interpretation of the change 
in the (super-) CM part of the angular momentum in a stationary-to-stationary
transition.
The first term $\Delta k_{\vec Y}$ represents a change in the (super-) CM 
angular momentum, which is similar to the integral of the flux of the 
intrinsic angular momentum (but involves the electric-parity part of the 
integrand, rather than the magnetic-parity part).
The last term in Eq.~\eqref{eq:DeltaKYu1u2} represents the change in the CM 
part of the angular momentum that arises from integrating the time dependence 
of a term like the supermomentum associated with the quantity $D_A Y^A$.
This term would typically grow linearly with $u$ when there is a net change 
in the supermomentum; however, the middle term in Eq.~\eqref{eq:DeltaKYu1u2}
also grows linearly with $u$, and will cancel this growth from the last term. 
The quantity $\Delta K_{\vec \zeta_Y}(u_1,u_2)$, therefore, is finite for 
spacetimes that radiate supermomentum over finite retarded-time intervals 
$u\in[u_1,u_2]$, and it contains information about the time dependence 
of the supermomentum beyond what is given by the net change in the 
supermomentum.\footnote{\label{fn:limits}~If we take
the limits $u_1\rightarrow -\infty$ and $u_2\rightarrow +\infty$, then we 
must make additional assumptions about the rate at which the supermomentum 
approaches a constant in the limits $u\rightarrow \pm\infty$ to ensure that
the change in CM angular momentum is finite.
For example, if we assume the leading-order time dependence goes as
$P_{(D_A Y^A)} \sim P_0 (1 + |u/u_0|^{-n})$ as $u\rightarrow \pm\infty$
($u_0$ is a reference time), then it is clear that we would need
to require $n>1$.
A detailed study of these types of asymptotics is beyond the scope of this
work, and it will not be necessary for spacetimes that radiate for a finite
interval of retarded time, $u\in[u_1,u_2]$.}
Thus, although the (super) CM part of the angular momentum can be made to 
vanish in a stationary region, its change in a stationary-to-stationary 
transition does not necessarily vanish.
In addition, it contains additional information that is not captured in the 
net changes in the supermomenta or in the other BMS charges.

There is a interesting feature specific to the flux of (super-) CM angular
momentum that we now point out.
Suppose we specialize Eq.~\eqref{eq:FluxChange} to the case in which 
$\alpha = u_0$ is a constant shift in retarded time.
All terms except the first vanish, and we find that
\begin{equation}
\Delta K_{\vec \zeta_Y}(u_2',u_1') - \Delta K_{\vec \zeta_Y}(u_2,u_1) 
= -\frac{u_0}2 \Delta P_{(D_A Y^A)} \, .
\label{eq:DeltaKu0}
\end{equation}
Thus, when there is a net change in the linear momentum or the supermomentum,
the change in the (super) CM angular momentum is not invariant under shifts in 
the cuts by constant values of $u_0$.
This transformation property of the CM angular momentum could be useful for
defining a specific BMS frame in an asymptotically flat spacetime.
As we noted in the previous subsection, the canonical frame associated with
the initial stationary region fixes all the BMS transformations except for a 
time translation $\vec\zeta = u_0\vec\partial_u$ and a global SO(3) rotation.
In these stationary regions, the charges are invariant under time translations;
thus, they cannot be used to determine a ``preferred'' retarded time in a 
stationary region.  
When there is a flux of 4-momentum (i.e., when $\Delta P_{(D_A Y^A)}\neq 0$), 
its time dependence allows for a preferred reference time (i.e., 
an ``origin'' of the time coordinate) to be picked out.
One natural choice comes from requiring that the magnitude of the change in 
the CM angular momentum  be minimized.
This is satisfied by a value of $u_0$ given by
\begin{equation}
u_0 =\frac{2 \Delta K_{\vec \zeta_Y} \Delta P_{(D_A Y^A)} }
{ \Delta P_{(D_A Y^A)}^2} \, . 
\label{eq:u0minDeltaK}
\end{equation}
This value of $u_0$ can be computed from changes in the BMS charges at 
infinity, and it is a geometrically motivated method of determining a 
reference time for spacetimes that radiate 4-momentum.
A possible application of this property of the CM angular momentum is defining 
a reference time for comparing gravitational waveforms from numerical 
relativity simulations of compact binaries that radiate linear momentum.
While we will not investigate this point in greater detail in this paper,
we will make use of this reference time for computing the change in the 
CM angular momentum in Sec.~\ref{sec:CMpn}.

\subsection{Multipole expansion of the flux of CM angular momentum carried
by GWs}

To compute an expression for the flux of the CM angular momentum carried by GWs
in terms of a set of multipole moments of the GW strain, we will closely 
follow the methods used to calculate the GW memory and spin memory effects 
given in~\cite{Nichols:2017rqr}.
We will expand $C_{AB}$ in terms of electric- and magnetic-parity
tensor spherical harmonics as
\begin{equation}
C_{AB} = \sum_{l,m} (U_{lm} T_{AB}^{(e),lm} + V_{lm} T_{AB}^{(b),lm}) \, ,
\label{eq:CABmultipoles}
\end{equation}
where the conventions we use for the second-rank tensor spherical harmonics
are given in an appendix of~\cite{Nichols:2017rqr}.
Because the tensor $C_{AB}$ is real, and because the tensor spherical harmonics
satisfy the relationships
\begin{equation}
T_{AB}^{(e),l-m} = (-1)^m \bar T_{AB}^{(e),lm} \, , \qquad
T_{AB}^{(b),l-m} = (-1)^m \bar T_{AB}^{(b),lm}
\end{equation}
(where the overline denotes complex conjugation), the coefficients of this 
expansion in spherical harmonics obey the related properties
\begin{equation}
U_{l-m} = (-1)^m \bar U_{lm} \, , \qquad
V_{l-m} = (-1)^m \bar V_{lm} \, .
\label{eq:UlmVlmConj}
\end{equation}
These tensor spherical harmonics are also related to spin-weighted spherical
harmonics, a complex vector 
\begin{equation}
\vec m = \frac 1{\sqrt 2}
(\vec \partial_\theta + i \csc\theta \vec \partial_\phi) \, ,
\label{eq:mVecDef}
\end{equation}
and its complex conjugate.
The GW flux in Eq.~\eqref{eq:FluxBondi} (which is a product of the shear, 
the news tensor, and the derivatives of both quantities) can be expressed as a
product of vector and second- and third-rank tensor spherical harmonics
(see~\cite{Nichols:2017rqr} for more detail).
For simplicity, we will assume that the stress-energy tensor of the matter 
fields vanishes, although this could be included trivially, because the flux
is linear in the material stress-energy tensor.

To compute the $l=1$ moments of the flux, we will integrate 
minus\footnote{Because GWs carry away energy from an isolated system with no 
incoming radiation, the flux is always negative.
Thus, it has become a common convention (e.g.,~\cite{Thorne1980}) to define
the energy carried away by GWs as a positive number, with it being implicit
that this positive change in the energy causes the Bondi mass of 
the system to decrease.
Similar sign conventions are used for the linear momentum and intrinsic part
of the angular momentum.
We also follow this convention with the flux of CM angular momentum, but
we add the superscript ``(GW)'' to this flux to make this convention explicit.}
the flux in Eq.~\eqref{eq:FluxBondi} against vector fields of the form 
$Y_A = D_A \bar Y_{1,m}$, where $Y_{l,m}(\theta,\phi)$ are scalar spherical
harmonics with the Condon-Shortley phase convention.
It is then possible to express the multipole moments of the flux in terms of 
integrals of products of three spin-weighted spherical harmonics (with the 
conventions for the harmonics given in~\cite{Nichols:2017rqr}).
Before evaluating these integrals, it will again be useful to perform 
integration by parts on the set of terms in Eq.~\eqref{eq:FluxBondi} that are 
the divergence of a scalar quantity.
Once this is done, the flux splits naturally into two types of terms 
\begin{equation}
\frac{d K_{1,m}^{(\mathrm{GW})}}{du} = \frac{d k_{1,m}^{(\mathrm{GW})}}{du} 
- u \frac{d P_{1,m}^{(\mathrm{GW})}}{du} \, ,
\label{eq:CMfluxSplit}
\end{equation}
as in Eq.~\eqref{eq:dKduSplit}.
Note that the apparent factor of $-2$ difference between the second 
terms on the right-hand sides of Eqs.~\eqref{eq:dKduSplit} 
and~\eqref{eq:CMfluxSplit} comes from a difference in convention for the 
supermomentum associated with a scalar function $D_A Y^A$ and the convention 
commonly used for the $l=1$ moments of the flux of linear momentum.
The multipolar expansion of the first term on the right-hand side of 
Eq.~\eqref{eq:CMfluxSplit} has not been computed before (as far as we are 
aware).
The second term is the same as the flux of linear momentum multiplied by 
minus the retarded time $u$.
The multipolar expansion of the linear-momentum flux has been computed before 
(for example, in~\cite{Thorne1980}).

The integrals of products of three spin-weighted spherical harmonics that 
arise in the flux of the CM angular momentum are relatively simple functions
of $l$ and $m$.
It will be helpful to define a few coefficients, so as to express the 
multipolar expansion for the CM angular momentum flux produced by GWs 
more concisely.
These coefficients are
\begin{subequations}
\begin{align}
a_l = & \sqrt{\frac{(l-1)(l+3)}{(2l+1)(2l+3)} } \, ,\\
b_{lm}^{(\pm)} = & \sqrt{(l\pm m +1)(l \pm m + 2)} \, , \\
c_{lm} = & \sqrt{(l - m +1)(l + m + 1)} \, , \\
d_{lm}^{(\pm)} = & \sqrt{(l\pm m + 1)(l\mp m)} \, .
\end{align}
\end{subequations}
After a lengthy calculation, it is possible to write the first term on the 
right-hand side of Eq.~\eqref{eq:CMfluxSplit} as
\begin{subequations}
\begin{align}
\frac{dk_{1,0}^{(\mathrm{GW})}}{du} = & -\frac{1}{64\pi} \sqrt{\frac 3\pi}
\sum_{l,m} a_l c_{lm} [\bar U_{lm} \dot U_{(l+1)m} \nonumber \\
& - \bar U_{(l+1)m} \dot U_{lm} + \bar V_{lm} \dot V_{(l+1)m} 
\nonumber \\
& - \bar V_{(l+1)m} \dot V_{lm} ] \, , \\
\frac{dk_{1,\pm 1}^{(\mathrm{GW})}}{du} = & -\frac 1{64\pi} \sqrt{\frac 3{2\pi}}
\sum_{l,m} a_l [b_{lm}^{(\pm)}(\bar U_{lm} \dot U_{(l+1)m\pm 1} 
\nonumber \\
& + \bar V_{lm} \dot V_{(l+1)m\pm 1}) - b_{lm}^{(\mp)}(\bar U_{(l+1)m\mp 1}
\dot U_{lm} \nonumber \\
& +\bar V_{(l+1)m\mp 1} \dot V_{lm})] \, , 
\end{align}
\label{eq:dkGW1mdu}%
\end{subequations}
and the second term as
\begin{subequations}
\begin{align}
\frac{dP_{1,0}^{(\mathrm{GW})}}{du} = & \frac{1}{32\pi} \sqrt{\frac 3\pi} 
\sum_{l,m} \frac {1}{l+1} [a_l c_{lm} (\dot{\bar U}_{lm} \dot U_{(l+1)m} 
\nonumber \\
& + \dot{\bar V}_{lm} \dot V_{(l+1)m}) - \frac{2im}{l} 
\dot{\bar U}_{lm} \dot V_{lm} ] \, , \\
\frac{dP_{1,\pm 1}^{(\mathrm{GW})}}{du} = & \frac{1}{32\pi} \sqrt{\frac 3{2\pi}}
\sum_{l,m} \frac{1}{l+1} [a_l b_{lm}^{(\pm)} (\dot{\bar U}_{lm} 
\dot U_{(l+1)m\pm 1} \nonumber \\
& + \dot{\bar V}_{lm} \dot V_{(l+1)m\pm 1}) \pm \frac{2i}{l} d_{lm}^{(\pm)} 
\dot{\bar U}_{lm} \dot V_{l(m\pm 1)} ] \, .
\end{align}
\label{eq:dPGW1mdu}%
\end{subequations}
All the sums in Eqs.~\eqref{eq:dkGW1mdu} and~\eqref{eq:dPGW1mdu} run over 
$l\geq 2$, and $-l\leq m \leq l$.

The translation subgroup of the BMS group is four-dimensional, and it can be 
treated as a manifold with a flat Minkowski metric
(see, e.g.,~\cite{Geroch1977}).
We can then express the $l=1$ moments of the flux of CM angular momentum in 
terms of vectors on this flat Minkowski manifold. 
Here, we will give the components in a set of Cartesian-type coordinates,
$(x,y,z)$, which we define from the spherical-polar coordinates $(\theta,\phi)$
commonly used with the spherical harmonics employed in this paper.
A method to transform from the $l=1$ moments to the Cartesian components 
is described in~\cite{Flanagan:2015pxa}, which we now summarize.

First, define the unit vector $n^i$ and its gradient with respect to the
derivative operator $D_A$ via
\begin{equation}
n^i = (\sin\theta\cos\phi, \sin\theta\sin\phi, \cos\theta) \, , \qquad
e_A^i = D_A n^i \, .
\label{eq:eAiDef}
\end{equation}
The 1-form $D_A \bar Y_{1,m}$ can then be expressed in terms of a linear 
combination of the Cartesian components of $e_A^i$ as
\begin{equation}
D_A \bar Y_{1,m} = \omega_{0i}^{1,m} e_A^i \, ,
\end{equation}
where the coefficients $\omega_{0i}^{1,m}$ are given by
\begin{subequations}
\begin{align}
& \omega_{0x}^{1,0} = 0 = \omega_{0y}^{1,0} \, , \qquad \omega_{0z}^{1,0} = 
\frac 12 \sqrt{\frac 3\pi} \, \\
& \omega_{0x}^{1,\pm 1} = \mp \frac 12 \sqrt{\frac 3{2\pi}} \, , \qquad
\omega_{0y}^{1,\pm 1} = \frac i2 \sqrt{\frac 3{2\pi}} \, , \qquad
\omega_{0z}^{1,\pm 1} = 0 \, .
\end{align}
\end{subequations}
Then, from the fact that the flux of CM angular momentum satisfies
\begin{equation}
\frac{dK^{(\mathrm{GW})}_{1,m}}{du} = \omega_{0i} 
\frac{dK^i_{(\mathrm{GW})}}{du} \, ,
\end{equation}
we find that the Cartesian components are
\begin{subequations}
\begin{align}
\frac{dK^{x}_{(\mathrm{GW})}}{du} = & -\sqrt{\frac{2\pi} 3} \Bigg[ \left( 
\frac{dk_{1,1}^{(\mathrm{GW})}}{du} - \frac{dk_{1,-1}^{(\mathrm{GW})}}{du}
\right) \nonumber \\
& -u \left( \frac{dP_{1,1}^{(\mathrm{GW})}}{du} 
- \frac{dP_{1,-1}^{(\mathrm{GW})}}{du}\right) \Bigg] \, , \\
\frac{dK^{y}_{(\mathrm{GW})}}{du} = & -i \sqrt{\frac{2\pi} 3} \Bigg[ \left( 
\frac{dk_{1,1}^{(\mathrm{GW})}}{du} + \frac{dk_{1,-1}^{(\mathrm{GW})}}{du} 
\right)  \nonumber \\
& - u \left( \frac{dP_{1,1}^{(\mathrm{GW})}}{du} 
+ \frac{dP_{1,-1}^{(\mathrm{GW})}}{du} \right) \Bigg] \, ,\\
\frac{dK^{z}_{(\mathrm{GW})}}{du} = & 2\sqrt{\frac{\pi}3} \left( 
\frac{dk_{1,0}^{(\mathrm{GW})}}{du} - u \frac{dP_{1,0}^{(\mathrm{GW})}}{du} 
\right) \, .
\end{align}
\label{eq:KModesToCartesian}%
\end{subequations}
As was explained in more detail in the previous subsection, the flux in 
Eq.~\eqref{eq:KModesToCartesian} represents the change in the CM part of the
angular momentum, for which the origin of the retarded time coordinate is 
chosen to be $u=0$.
It contains nontrivial information about the flux of CM angular momentum from
the system that is not contained in the fluxes of the other BMS charges.

\section{\label{sec:CMmemory} Center-of-mass gravitational-wave memory effect}

In this section, after giving an argument for why the CM memory effect should
exist, we define the effect, describe some of its basic properties, and derive 
an expansion for the CM memory effect in terms of multipole moments of the 
GW strain.

\subsection{Rationale for the existence of the CM memory effect}

Consider, for simplicity, an asymptotically flat spacetime undergoing a
stationary-to-stationary transition as it radiates GWs for a finite time.
In each stationary region, there is a canonical reference frame in which
the Bondi mass aspect is constant, the shear vanishes, and the Bondi 
angular-momentum aspect is a linear combination of $l=1$ magnetic-parity
vector spherical harmonics (though the values of the mass and 
angular-momentum aspects will generally be different in the canonical frames 
of the two stationary regions).
The two canonical frames typically will not be the same, but there will be a
BMS transformation (a Lorentz transformation and supertranslation) that 
relates the two.
The supertranslation between the two canonical frames is equivalent to the
GW memory (e.g.,~\cite{Flanagan:2015pxa}), and the Lorentz transformation 
is related to the change in the 4-momentum between and the relative 
rotation of a set of fiducial observers in each of the two stationary regions.

Next, we integrate Eq.~\eqref{eq:FluxBondi} with respect to $u$ to 
relate the change in the charges to the net flux between the cuts:
\begin{align}
\Delta K_{\vec \zeta_Y} = & -\frac 1{64\pi} \int_{u_1}^{u_2} du \int d^2\Omega 
Y^A [u D_A ( 2D_B D_C N^{BC} \nonumber \\ 
& - N_{BC} N^{BC} - 32\pi \hat T_{uu}) + C^{BC} D_B N_{AC} \nonumber \\
& - N^{BC} D_B C_{AC} + 3N_{AB} D_C C^{BC} \nonumber \\
& - 3C_{AB} D_C C^{BC} +64\pi\hat T_{uA} + 16\pi \partial_u \hat T_{rA}] \, .
\label{eq:CMchargeChange}
\end{align}
The left-hand side of Eq.~\eqref{eq:CMchargeChange}, the change in the 
charges, depends on just the values of the 4-momentum and angular momentum in 
the canonical frames in the stationary regions and the BMS transformation that 
contains information about the net rotation, boost, and supertranslation 
between the two canonical frames.
We argued in Sec.~\ref{sec:CMflux}, however, that the net change in the 
(super-) CM angular momentum, as computed using the right-hand side of 
Eq.~\eqref{eq:CMchargeChange}, contains additional information besides the 
change in the supermomentum, angular momentum, and GW memory.
Thus, there appears to be an inconsistency.
It could be resolved, if there is a cancellation between certain terms in 
the flux, for example.

Such a cancellation occurs with the GW memory, which we will now review. 
First, recall that the potential $\Delta\Phi$ that determines the memory 
[see Eq.~\eqref{eq:MemDeltaPhi}] can be found by integrating the 
conservation-type equation for the Bondi mass aspect~\eqref{eq:Aspects}:
\begin{equation}
\mathcal D \Delta \Phi = \mathcal P \left [8\Delta m 
+ \int_{u_1}^{u_2} du( 32\pi \hat T_{uu} + N_{AB} N^{AB} ) \right]
\label{eq:DeltaPhiMem}
\end{equation}
(see, e.g.,~\cite{Flanagan:2015pxa}).
We have defined a differential operator 
\begin{equation}
\mathcal D \equiv D^2(D^2+2) 
\end{equation}
and a projector $\mathcal P$, which removes the $l=0$ and $l=1$ spherical 
harmonics from the right-hand side of Eq.~\eqref{eq:DeltaPhiMem}.
The projector is needed to invert the operator $\mathcal D$ and solve for 
$\Delta\Phi$, because the $l=0$ and $l=1$ harmonics are in its kernel of
the operator $\mathcal D$.
In the terminology of~\cite{Bieri:2013ada}, the first term on the right-hand
side in Eq.~\eqref{eq:DeltaPhiMem} is the ordinary part of the GW memory, and 
the remaining terms in the integral are collectively the null part of the 
memory.
We will, therefore, express $\Delta \Phi$ as a sum of two parts
\begin{equation}
\Delta \Phi = \Delta \Phi_{(\mathrm o)} + \Delta \Phi_{(\mathrm n)} \, ,
\end{equation}
which correspond to the parts of the solution to Eq.~\eqref{eq:DeltaPhiMem}
for the ordinary and null parts, respectively (and which is possible because
the equation is linear in $\Delta\Phi$).

Now, let us return to the cancellation that occurs for the GW memory effect.
The ordinary part of the memory (the change in the supermomentum charges, up
to a normalization factor) depends on just the net change in the rest mass of
the system and the relative boost of the observers that determine the 
canonical frames. 
The null memory, however, can be arbitrary. 
Thus, the integral of $D_A D_B N^{AB}$ with respect to $u$ must be nonzero, so
that Einstein's equations (and, equivalently, charge conservation) are 
satisfied.
For the spin memory, the values of the change in the superspin charges are
also restricted in a stationary-to-stationary transition, but the null part of 
the spin memory is not limited in this way.
The additional term in the flux, Eq.~\eqref{eq:FluxSMterm}, therefore, is 
necessary to ensure that charge conservation holds.

Finally, let us revisit the ``inconsistency'' discussed below 
Eq.~\eqref{eq:CMchargeChange} about the change in the charges in light of 
the discussion above.
The related inconsistencies for supermomentum and superspin charge conservation 
were resolved by the GW memory and spin memory effects, respectively.
Thus, it seems natural to suggest a similar resolution for super-CM charge
conservation: namely, that there must be a CM memory effect.
Because the GW memory and spin memory effects come about from terms in
the fluxes that are linear in the Bondi news and shear tensors, respectively, 
we expect that the CM memory will arise from a similar type of term in the
flux of CM angular momentum.
The term linear in the Bondi news tensor in Eq.~\eqref{eq:CMchargeChange} 
(proportional to $u D_A D_B D_C N^{BC}$), therefore, is the most obvious term 
that could give rise to the CM memory.
As we discuss in the next subsection, it turns out to be the $u$ integral of 
a quantity related to this term that will be the CM memory effect.

\subsection{Definition and properties of the CM memory effect}

Let us then define a quantity
\begin{align}
\Delta \mathcal C_{(D_A Y^A)} \equiv & -\int_{u_1}^{u_2} \! du \int d^2\Omega 
\, u \left(D_B D_C N^{BC} - \frac 12 \mathcal D \dot \Phi_{(\mathrm n)} 
\right) \nonumber \\
& \times (D_A Y^A) \, ,
\label{eq:CMmemDef}
\end{align}
which should be interpreted as a part of $u$ times the $u$ integral of a
quantity proportional to a portion of the Bondi news tensor, with the part of 
the news tensor responsible for the null GW memory removed [this latter part of 
the news tensor is denoted by the potential 
$\dot\Phi_{(\mathrm n)}$].\footnote{Using Einstein's 
equations~\eqref{eq:Aspects}, we could have written this as a term 
proportional to $u$ times the $u$ derivative of the Bondi mass 
aspect.
We deliberately avoided writing it in this form, so as to reinforce the 
notion that this is an observable with the units of the time integral of the 
GW strain, rather than a quantity with the units of the time integral of 
the supermomentum.}
This quantity has the units of the time integral of the GW strain (like the
spin memory effect), and it will be our definition of the CM memory effect.
We now investigate some of its properties.

Integrating Eq.~\eqref{eq:CMmemDef} by parts with respect to $u$, we find that
\begin{align}
\Delta \mathcal C_{(D_A Y^A)} = & \int_{u_1}^{u_2} du \int d^2\Omega
\left( D_B D_C C^{BC} -\frac 12 \mathcal D \Phi_{(\mathrm n)} \right)
\nonumber \\
& \times (D_A Y^A) - u \int d^2\Omega (D_A Y^A) \nonumber \\
& \times \left.\left(D_B D_C C^{BC} 
-\frac 12 \mathcal D \Phi_{(\mathrm n)} \right) \right|^{u_2}_{u_1} \, .
\end{align}
Thus, we see that $\Delta \mathcal C_{(D_A Y^A)}$ contains information about 
the time integral of $C_{AB}$, but it removes the part that grows linearly 
with $u$, which arises when there is ordinary GW memory.\footnote{A similar
caveat to that elaborated in footnote~\ref{fn:limits} holds: namely, for finite 
values of $u_1$ and $u_2$, we do not need to suppose that the ordinary GW 
memory approaches a constant at a given rate; however, in the limit that 
$u_1\rightarrow -\infty$ and $u_2\rightarrow +\infty$, we would need to 
assume similar fall-off rates to those given for the supermomentum in
footnote~\ref{fn:limits}.}
It is, therefore, the part of the time integral of the electric-parity part of 
$C_{AB}$ that becomes constant in a stationary-to-stationary transition.

Next, we will consider how the CM memory effect behaves in a set of cuts
that are supertranslated from the cuts $u$ used to compute the effect above.
Under a supertranslation, $\alpha$, the news tensor transforms as
\begin{equation}
\delta N_{AB} = \alpha \dot N_{AB} \, ,
\end{equation}
to linear order in $\alpha$.
Using this relationship, integration by parts, and the facts that 
$u'=u+\alpha$ and the news tensor and $\hat T_{uu}$ vanish in a nonradiative 
region, it is then straightforward to show from Eq.~\eqref{eq:CMmemDef} that 
\begin{align}
\Delta \mathcal C'_{(D_A Y^A)} = & -\int_{u'_1}^{u'_2} du' \int d^2\Omega 
u' \, \bigg(D^B D^C N_{BC}' \nonumber \\
& - \frac 12 \mathcal D \dot \Phi_{(\mathrm n)}' \bigg)
(D_A Y^A) = \Delta \mathcal C_{(D_A Y^A)} \, .
\end{align}
In the equation above, we computed $\Delta \mathcal C'_{(D_A Y^A)}$ with 
respect to the generators adapted to cuts of constant $u'$.
Thus, $\Delta \mathcal C_{(D_A Y^A)}$ is invariant under infinitesimal 
supertranslations, for stationary-to-stationary transitions.

When computing memory effects within the Bondi framework, it can be
useful to define a scalar potential as the memory observable
(see, e.g.,~\cite{Flanagan:2015pxa}). 
We now define this quantity.
The shear tensor, $C_{AB}$, can be expressed in terms of two potentials that
encompass its two degrees of freedom as follows:
\begin{equation}
C_{AB} = \frac 12 (2D_A D_B - h_{AB} D^2) \Phi 
+ \epsilon_{C(A} D_{B)} D^C \Psi \, .
\label{eq:CABdecomp}
\end{equation}
Using this decomposition and integrating by parts with respect to $u$, we find 
that Eq.~\eqref{eq:CMmemDef} can be written as 
\begin{align}
\Delta \mathcal C_{(D_A Y^A)} = & \frac 12 \int d^2\Omega (D_A Y^A) \mathcal D
\bigg[\int_{u_1}^{u_2} du(\Phi - \Phi_{(\mathrm n)}) \nonumber \\
& - u (\Phi - \Phi_{(\mathrm n)}) |_{u_1}^{u_2} \bigg] \, .
\end{align}
The CM memory observable that we define is 
\begin{equation}
\Delta \mathcal K \equiv \int_{u_1}^{u_2} du \, 
(\Phi - \Phi_{(\mathrm n)}) 
- u(\Phi - \Phi_{(\mathrm n)}) |_{u_1}^{u_2} \, ,
\label{eq:DeltaCalK}
\end{equation}
which is a potential for the time integral of the electric-parity part of the
shear with the part that grows linearly with $u$ from the ordinary part of the 
GW memory removed.\footnote{Given the somewhat complicated nature of the
CM memory observable, the reader might be concerned about whether this quantity
is measurable by freely falling observers, in principle.
Because the GW strain, the GW memory, and their time integrals can be measured
by freely falling observers, the basic ingredients needed to construct the
CM memory observable are measurable.
The CM memory effect corresponds to the electric-parity part of the 
time-integrated GW strain, and this part could be separated from the 
magnetic-parity part by having many observers surrounding an isolated source
measuring the GW strain.
Thus, the one remaining potential subtlety relates to extracting just the null
part of the memory.
This could be performed by directly measuring the flux of GWs and 
massless fields with appropriate detectors or by determining the time
dependence of the ordinary part of the memory (i.e., the flux of the
supermomentum charges) by measuring components of the asymptotic Riemann 
tensor with a generalization of the procedure described 
in~\cite{Flanagan:2014kfa,Flanagan:2016oks}, for example.
Thus, we see no obstacle for observing the CM memory, in principle, but a 
more detailed analysis of its measurability would be beneficial.}
The quantity $\Delta \mathcal C_{(D_A Y^A)}$ can be expressed in terms of 
$\Delta \mathcal K$ by
\begin{equation}
\Delta \mathcal C_{(D_A Y^A)} = \frac 12 \int d^2\Omega (D_A Y^A) \mathcal D
\Delta \mathcal K \, .
\end{equation}

Using Eq.~\eqref{eq:CMchargeChange}, we can also solve for 
$\Delta \mathcal C_{(D_A Y^A)}$ from the change in the super-CM angular 
momentum and the quadrupole and higher multipole moments of the flux of CM 
angular momentum carried by GWs and matter fields:
\begin{align}
\Delta \mathcal C_{(D_A Y^A)} = & -32 \pi \mathcal P \Delta K_{\vec \zeta_Y} 
- \frac 12 \mathcal P \int_{u_1}^{u_2} du \int d^2\Omega Y^A \nonumber \\
& \times (C^{BC} D_B N_{AC} - N^{BC} D_B C_{AC} \nonumber \\
& + 3N_{AB} D_C C^{BC} - 3C_{AB} D_C C^{BC} \nonumber \\
& +64\pi\hat T_{uA} + 16\pi \partial_u \hat T_{rA}) \, .
\label{eq:CMmemFlux}
\end{align}
As with the GW memory and spin memory, the CM memory has two parts: the first
given by $\Delta K_{\vec \zeta_Y}$ is the ordinary part, whereas the portion
involving the retarded-time integral is the null part.
Because the CM memory is invariant under infinitesimal supertranslations, but 
the changes in the CM part of the super angular momentum transform in the 
way given in Eq.~\eqref{eq:FluxChange}, then the ordinary and null parts of 
the CM memory must transform in opposite ways.

To more easily compute the amplitude of the CM memory effect produced by 
astrophysical sources, we expand Eq.~\eqref{eq:CMmemFlux} in a set of
multipole moments of the GW strain in the next subsection.

\subsection{Multipolar expansion of the CM memory effect}

We find it simplest to compute the multipole moments of the CM memory effect
by integrating the right-hand side of Eq.~\eqref{eq:CMmemFlux} with respect 
to the smooth vector fields.
Specifically, we use the electric-parity vector spherical harmonics, 
$D_A \bar Y_{lm} /\sqrt{l(l+1)}$ (analogously to what was done in the 
calculation of the spin memory in~\cite{Nichols:2017rqr}).
These functions are a useful basis for smooth vector fields, like those 
used by Campiglia and Laddha~\cite{Campiglia:2014yka,Campiglia:2015yka}.
Because the integral of a meromorphic super-rotation vector 
field~\cite{Barnich2009} with a smooth vector field is finite 
(see, e.g.,~\cite{Flanagan:2015pxa}), then it could also represent the 
part of the super-rotation symmetry that has overlap with these vector 
spherical harmonics (although this decomposition may not be 
unique~\cite{Compere:2016jwb}).

The method for calculating the multipole moments of the CM memory observable, 
which we will denote by $\Delta \mathcal K_{lm}$, is very similar to the 
procedure to compute similar moments of the spin memory described 
in~\cite{Nichols:2017rqr} (as well as that described in Sec.~\ref{sec:CMflux}
for computing the multipole moments of the CM angular-momentum flux).
The basic strategy of the calculation is to change the tensorial expression
for the multipole moments of the shear and its derivatives into a sum of 
products of three spin-weighted spherical harmonics.
The conventions for the vector, tensor, and spin-weighted harmonics are 
given in detail in~\cite{Nichols:2017rqr}.
Using these conventions, we define a set of coefficients
\begin{align}
& \mathcal B_l(s',l',m';s'',l'',m'') \equiv \nonumber \\
& \int d^2\Omega (_{s'}Y_{l'm'})(_{s''}Y_{l''m''})
(_{s'+s''}\bar Y_{l(m'+m'')}) \, ,
\end{align}
as in~\cite{Nichols:2017rqr}.\footnote{The coefficients 
$\mathcal B_l(s',l',m';s'',l'',m'')$ are identical to the coefficients 
denoted $\mathcal C_l(s',l',m';s'',l'',m'')$ in~\cite{Nichols:2017rqr}; 
however, we have renamed them here, so as to avoid confusion with the quantity
$\Delta \mathcal C_{(D_A Y^A)}$ that is related to the CM memory effect.}
We have restricted to integrals in which the complex-conjugated spin-weighted
spherical harmonic has spin weight $s=s'+s''$ and has azimuthal number 
$m=m'+m''$, because the integrals are zero for all other values of $s$ and $m$.
Furthermore, the only values of $l$ for which the integral is nonvanishing
are those with 
$l\in\{\max(|l'-l''|,|m'+m''|,|s'+s''|),\ldots,l'+l''-1,l'+l'' \}$.
The reason for these ``selection rules'' comes from the fact that the 
coefficients $\mathcal B_l(s',l',m';s'',l'',m'')$ can be expressed in terms 
of products of Clebsch-Gordan coefficients 
$\langle l',m';l'',m''|l,m'+m''\rangle$ via the relationship
\begin{align}
& \mathcal B_l(s',l',m';s'',l'',m'') \nonumber \\
& \quad = (-1)^{l+l'+l''} \sqrt{\frac{(2l'+1)(2l''+1)}{4\pi(2l+1)}} 
\nonumber \\
& \quad \times \langle l',s';l'',s''|l,s'+s''\rangle 
\langle l',m';l'',m''|l,m'+m''\rangle \, .
\end{align}
These coefficients, therefore, satisfy similar identities to those of the
Clebsch-Gordan coefficients when the signs of the spin weight or the 
azimuthal numbers are changed (see, e.g.,~\cite{Nichols:2017rqr}).

Next, we specialize to vacuum spacetimes, and we compute the multipole moments 
$\Delta \mathcal K_{lm}$ of the CM memory produced by GWs.
Nonvacuum cases can be treated by simply adding the appropriate multipole
moments of the relevant components of the stress-energy tensor given in 
Eq.~\eqref{eq:CMmemFlux}.
To make the expression more compact, we write the result as
\begin{align}
\Delta \mathcal K_{lm} = & \frac{(l-2)!}{(l+2)!} \frac 1{\sqrt{l(l+1)}} 
\mathcal P \Bigg( \int_{u_1}^{u_2} du \frac{dk^{(\mathrm{CM})}_{lm}}{du}
\nonumber \\
& + 64\pi \Delta K_{lm} \Bigg) \, .
\label{eq:DeltaKlm}
\end{align}
The first term in the integral $dk^{(\mathrm{CM})}_{lm}/du$ comes from the
higher multipole moments of the quantity that gives rise to the term 
$dk^{(\mathrm{GW})}_{1m}/du$ in the flux of CM angular momentum (though with 
a different overall normalization).
The second term, $\Delta K_{lm}$, is a spherical harmonic moment of the 
change in the super-CM charges $\Delta K_{\vec \zeta_Y}$.
Before giving the explicit form of the term $dk^{(\mathrm{CM})}_{lm}/du$, we 
make a few additional definitions of coefficients so as to write the 
result more compactly:
\begin{subequations}
\begin{align}
s^{l,(\pm)}_{l';l''} = & 1 \pm (-1)^{l+l'+l''} \, , \\
c^l_{l',m';l'',m''} = & 3\sqrt{(l'-1)(l'+2)}\mathcal B_l(-1,l',m';2,l'',m'')
\nonumber \\
& + \sqrt{(l''-2)(l''+3)} \mathcal B_l(-2,l',m';3,l'',m'')  \, .
\end{align}
\end{subequations}
After a lengthy calculation, it is possible to show that
\begin{align}
\frac{dk^{(\mathrm{CM})}_{lm}}{du} = & \frac{1}{4}
\sum_{l',l'',m',m''} c^l_{l',m';l'',m''} [
s^{l,(+)}_{l';l''} (U_{l'm'} \dot U_{l''m''} \nonumber \\
&  - \dot U_{l'm'} U_{l''m''} + V_{l'm'} \dot V_{l''m''} 
- \dot V_{l'm'} V_{l''m''})
\nonumber \\
& +i s^{l,(-)}_{l';l''} 
(U_{l'm'} \dot V_{l''m''} + \dot V_{l'm'}U_{l''m''} \nonumber \\
&  - \dot U_{l'm'} V_{l''m''} - V_{l'm'} \dot U_{l''m''}) ] \, .
\label{eq:dkCMlmdu}
\end{align}
The sum runs over $l',l'' \geq 2$, and for $-l' \leq m' \leq l'$ 
and $-l'' \leq m'' \leq l''$. 
For an arbitrary source, an infinite number of products of multipoles will be
needed to compute the CM memory effect.
For compact binaries in the PN approximation, the number of multipole moments
that contribute at leading order is a small number, as we discuss 
next.

\section{\label{sec:CMpn} Flux of CM angular momentum and CM memory in the 
PN approximation}

In this part, we introduce a few essential elements of the PN formalism for 
compact binaries that we need for the calculations in this section.
Our summary is based on the much more comprehensive 
review~\cite{Blanchet:2013haa}.
We then present the main results of this section: expressions for the 
leading-PN-order flux of CM angular momentum and CM memory effect for 
nonspinning, quasicircular compact binaries.
We also comment on the terms in the gravitational waveform responsible for
producing the CM memory effect and on the prospects for detecting these 
features in the waveform with future GW detectors.

\subsection{Summary of selected results from PN theory}

In PN theory, the gravitational waveform is typically described by 
a transverse-traceless tensor $h_{ij}^{\mathrm{TT}}$. 
It can be expanded in second-rank electric- and magnetic-parity tensor 
spherical harmonics as 
\begin{equation}
h_{ij}^{\mathrm{TT}} = \frac 1r \sum_{l,m} (U_{lm} T_{ij}^{(e),lm} + V_{lm} 
T_{ij}^{(b),lm}) \, ,
\label{eq:hijTT}
\end{equation}
where the sum runs over $l\geq 2$ and $-l\leq m \leq l$.
It was argued in~\cite{Nichols:2017rqr} that the coefficients $U_{lm}$ and
$V_{lm}$ that appear in both Eqs.~\eqref{eq:CABmultipoles} and~\eqref{eq:hijTT}
are the same in linearized theory (though care would need to be taken to 
properly include any nondynamical terms in $h_{ij}^{\mathrm{TT}}$, 
as noted by~\cite{Ashtekar:2017wgq}).
It is often convenient to work with the complex GW strain 
$h = h_+ - i h_\times$, 
which is related to the tensorial strains by
\begin{equation}
h = r^{-1} C_{AB} \bar m^A \bar m^B 
= h_{ij}^{\mathrm{TT}} e_A^i e_B^j \bar m^A \bar m^B \, ,
\end{equation}
where $m^A$ is defined in Eq.~\eqref{eq:mVecDef} and $e_A^i$ is given in 
Eq.~\eqref{eq:eAiDef}.
When $h$ is expanded in spin-weighted spherical harmonics, a short calculation
shows that
\begin{equation}
h = \sum_{l,m} h_{lm} (_{-2}Y_{lm}) \, , \qquad
h_{lm} = \frac 1{r\sqrt{2}} (U_{lm} - i V_{lm}) \, .
\end{equation}
The convention used here for $h$ differs from that in~\cite{Blanchet:2013haa} 
by an overall minus sign, but the multipole moments $U_{lm}$ and $V_{lm}$ as 
well as the tensorial GW strain agree (as they must).

Through a matching procedure, summarized in the review~\cite{Blanchet:2013haa},
it is possible to relate the radiative moments $U_{lm}$ and $V_{lm}$ to source 
multipole moments, $I_{lm}$ and $J_{lm}$ that, as their name suggests, 
describe the multipole moments of the source in the near zone.
To simplify the matching, we will choose our coordinates such that the orbital 
angular momentum of our nonspinning, compact-binary source points in the $z$ 
coordinate direction.
The matching procedure takes place through a third set of intermediate 
``canonical'' multipole moments,\footnote{While the term ``canonical'' is used
to describe both these moments and a specific Bondi frame associated with a 
stationary region in asymptotically flat spacetimes, this repeated usage is
just an unfortunate repetition of the term ``canonical''; there is no obvious
connection between the two concepts.} 
$M_{lm}$ and $S_{lm}$, as well as a set of multipole moments that parameterize 
a coordinate transformation between two solutions of the linearized Einstein's 
equations.
For multipoles with $m\neq0$, the relationship between the radiative and 
canonical moments is given by
\begin{equation}
U_{lm} = M_{lm}^{(l)} + O(c^{-3}) \, , \qquad 
V_{lm} = S_{lm}^{(l)} + O(c^{-3}) \, ,
\label{eq:RadToCanonical}
\end{equation}
where here---and everywhere else hereafter---the 
remainder means there are \textit{relative} PN corrections (where PN 
corrections conventionally scale as the power of $c$ to the minus one-half), 
and where the superscript $(l)$ means to take $l$ derivatives with respect to 
$u$.
These corrections consist of terms that get called ``tails'' (including 
higher PN-order generalizations, such as ``tails of tails''), ``instantaneous'' 
nonlinear terms, and ``hereditary'' (or ``memory'') terms.
For understanding the terms in the GWs that give rise to the CM memory effect, 
the instantaneous, nonlinear terms will play the most important role.
The canonical moments are related to the source moments by 
\begin{equation}
M_{lm} = I_{lm} + O(c^{-5}) \, , \qquad S_{lm} = J_{lm} + O(c^{-5}) \, .
\end{equation}
The 2.5PN remainder here means that there are additional nonlinear terms 
entering at this order in the PN expansion that are not captured by the PN 
expansion of the source multipoles $I_{lm}$ and $J_{lm}$ to that PN order.

For computing the leading-order flux of CM angular momentum and the CM memory 
effect, it turns out that we will be able to use the leading Newtonian 
expressions for the radiative multipole moments in terms of the source moments
(for $m\neq 0$),
\begin{equation}
U_{lm} = I_{lm}^{(l)} + O(c^{-3}) \, , \qquad 
V_{lm} = J_{lm}^{(l)} + O(c^{-3}) \, .
\label{eq:RadToSource}
\end{equation}
The $U_{2,0}$ mode below comes from the GW memory, which does not satisfy 
Eq.~\eqref{eq:RadToSource}, even though it is a leading, Newtonian-order 
effect in the waveform.
We will also need to use one higher-PN-order calculation for the 
flux of linear momentum carried by GWs.
For comparing the parts of the GWs responsible for the CM memory with the
expressions for the multipole moments of the waveform in PN theory, however, 
we will need to be aware of the higher-order PN corrections to the radiative
multipole moments.
Given that the corrections have distinct mathematical forms (tail, 
instantaneous, and hereditary terms), we will be able to identify the
relevant nonlinear terms to make this comparison analytically.
Identifying these terms observationally in the GWs from compact-binary 
mergers will be much more challenging.

For nonspinning compact binary sources in quasicircular orbits, the radiative 
multipole moments can be expressed conveniently in terms of just a few 
parameters, most of which involve the masses of the two bodies, $m_{(A)}$ 
and $m_{(B)}$: the total mass $M=m_{(A)}+m_{(B)}$, the mass difference 
$\delta m = m_{(A)} - m_{(B)}$, the symmetric mass ratio 
$\eta=m_{(A)} m_{(B)}/M^2$, the orbital frequency $\omega$, the PN parameter 
$x=(M\omega)^{2/3}$, and the orbital phase $\varphi$ 
(see, e.g.,~\cite{Blanchet:2013haa}).
In terms of these quantities, the radiative moments that we will need for our 
calculations are 
\begin{subequations}
\begin{align}
U_{2,2} = & - 8 \sqrt{\frac{2\pi}5} M \eta x e^{-i2\varphi} + O(c^{-2}) \, , \\
U_{2,0} = & \frac 47 \sqrt{\frac{5\pi}3} M \eta x + O(c^{-2}) \, , \\
U_{3,1} = & - \frac{2i}3 \sqrt{\frac{\pi}{35}} \delta m \eta x^{3/2} 
e^{-i\varphi} +  O(c^{-2}) \, , \\
U_{3,3} = & 6i \sqrt{\frac{3\pi}{7}} \delta m \eta x^{3/2} e^{-i3\varphi} 
+ O(c^{-2}) \, , \\
V_{2,1} = & \frac 83 \sqrt{\frac{2\pi}5} \delta m \eta x^{3/2} e^{-i\varphi} +
O(c^{-2}) \, ,
\end{align}
\label{eq:UlmPNvals}%
\end{subequations}
where the orbital phase is given by
\begin{equation}
\varphi = -\frac{x^{-5/2}}{32\eta} + O(c^{-2}) \, .
\end{equation}
The modes with negative azimuthal number can be obtained by using the 
relationships given in Eq.~\eqref{eq:UlmVlmConj}.
The $u$ derivatives of these multipole moments can be expressed in terms of
$x$ by using the chain rule and the fact that
\begin{equation}
\dot x = \frac{64\eta}{5M} x^5 + O(c^{-2}) \, .
\label{eq:xdot}
\end{equation}
The results are as follows:
\begin{subequations}
\begin{align}
\dot U_{2,2} = & 16 i \sqrt{\frac{2\pi}5} \eta x^{5/2} e^{-i2\varphi} 
+ O(c^{-2}) \, , \\
\dot U_{2,0} = & \frac{256}7 \sqrt{\frac{\pi}{15}} \eta^2 x^5 
+ O(c^{-2}) \, , \\
\dot U_{3,1} = & - \frac{2}3 \sqrt{\frac{\pi}{35}} \frac{\delta m}M \eta x^3 
e^{-i\varphi} +  O(c^{-2}) \, , \\
\dot U_{3,3} = & 18 \sqrt{\frac{3\pi}{7}} \frac{\delta m}M \eta x^3 
e^{-i3\varphi} + O(c^{-2}) \, , \\
\dot V_{2,1} = & -\frac{8i}3 \sqrt{\frac{2\pi}5} \frac{\delta m}M \eta x^3
e^{-i\varphi} + O(c^{-2}) \, .
\end{align}
\label{eq:UlmDotPNvals}%
\end{subequations}
Because the quantity $\dot U_{2,0}$ is several PN orders higher than the other
derivatives, it will not appear in most of the calculations below.

\subsection{Flux of CM angular momentum}

In this part, we give the leading-PN order expression for the flux of the CM 
part of the angular momentum.
We begin by computing the term $dk^{(\mathrm{GW})}_{1,1}/du$ in 
Eq.~\eqref{eq:CMfluxSplit}.
It is given by
\begin{align}
\frac{dk_{1,1}^{(\mathrm{GW})}}{du} = & -\frac 1{64\pi} \sqrt{\frac 3{7\pi}}
[\sqrt{15}(U_{2,-2} \dot U_{3,3} + U_{3,3}\dot U_{2,-2}) \nonumber \\
& + (U_{2,2} \dot U_{3,-1} + U_{3,-1}\dot U_{2,2}) 
+ \sqrt{6} U_{2,0} \dot U_{3,1} ] \nonumber \\
& + O(c^{-2}) \, .
\label{eq:dk11GWdu}
\end{align}
Substituting the relevant components in Eqs.~\eqref{eq:UlmPNvals} 
and~\eqref{eq:UlmDotPNvals} into Eq.~\eqref{eq:dk11GWdu}, we find that it
can be written as a function of $x$ as follows:
\begin{equation}
\frac{dk_{1,1}^{(\mathrm{GW})}}{du} = \frac{627}{980} \sqrt{\frac{3}{2\pi}}
\delta m \eta^2 x^4 e^{-i\varphi} + O(c^{-2}) \, .
\label{eq:dk11du}
\end{equation}
The $l=1$, $m=0$ term vanishes for nonspinning, quasicircular compact binaries.
This can be shown using arguments based on parity, like those given 
in~\cite{Boyle:2007sz}.

The second term on the right-hand side of Eq.~\eqref{eq:CMfluxSplit} requires
computing the flux of linear momentum.
This has been computed before (it can be inferred from~\cite{Damour:2006tr},
for example) and is given by 
\begin{align}
\frac{dP_{1,1}^{(\mathrm{GW})}}{du} = & \frac 1{96\pi} \sqrt{\frac 3{7\pi}}
(\sqrt{15} \dot U_{2,-2} \dot U_{3,3} + \dot U_{2,2} \dot U_{3,-1}
\nonumber \\ 
& +i\sqrt{14} \dot U_{2,2} \dot V_{2,-1} ) + O(c^{-2}) \, ,
\label{eq:dP11GWdu}
\end{align}
at leading PN order.
Inserting the appropriate values of the radiative moments given in 
Eq.~\eqref{eq:UlmDotPNvals} into Eq.~\eqref{eq:dP11GWdu}, we find
\begin{equation}
\frac{dP_{1,1}^{(\mathrm{GW})}}{du} = -i \frac{232}{105} \sqrt{\frac{3}{2\pi}}
\frac{\delta m}{M} \eta^2 x^{11/2} e^{-i\varphi} + O(c^{-2}) \, ,
\label{eq:dP11du}
\end{equation}
a result that traces back to~\cite{Fitchett1983}.
The $l=1$, $m=0$ mode of the flux of linear momentum also vanishes, which 
follows from the arguments based on parity in~\cite{Boyle:2007sz}.

To compute the net change in the CM angular momentum, we must evaluate
\begin{equation}
\Delta K_{1,1}^{(\mathrm{GW})} = \int du \left( 
\frac{dk_{1,1}^{(\mathrm{GW})}}{du} - u \frac{dP_{1,1}^{(\mathrm{GW})}}{du}
\right) \, .
\label{eq:DeltaK11}
\end{equation}
Because at leading order, the retarded time $u$ goes as
\begin{equation}
(u_c - u) = \frac{5M}{256\eta} x^{-4} + O(c^{-2})
\label{eq:uofx}
\end{equation}
(where $u_c$ is the retarded time of coalescence of the binary in PN theory),
then by comparing powers of $x$, we see that the first term on the right-hand
side of Eq.~\eqref{eq:DeltaK11} is 2.5 PN orders higher than the second term 
is.
While this might make the reader wonder why we do not neglect this term and 
focus just on the second term, we now revisit some of the discussion around 
Eqs.~\eqref{eq:DeltaKu0} and~\eqref{eq:u0minDeltaK} in the context of 
nonspinning PN compact binaries.

In PN theory, the BMS supertranslations are fixed by the fiducial Minkowski 
spacetime that is the background about which the PN expansion is computed.
This leaves the Poincar\'e group as the remaining symmetries.
There is a relatively natural way to fix the boost transformations (by moving
to the rest frame of the source in the initial stationary region, for example).
The rotations in the Lorentz group can be specified by aligning the orbital 
angular momentum to fall along the $z$ axis and the separation to be along 
the $x$ axis (at some fiducial time) in the initial stationary region.
Finally, one way to constrain the spatial translations is to require that
the CM of the system coincide with the origin of the coordinates initially.
This will make the CM part of the angular momentum equal to zero in this 
region.
A translation in time will not affect the values of the 4-momentum,
supermomentum, and (super) angular momentum in the initial stationary region
in this frame. 
Thus, there is no obvious prescription for using the (extended) BMS charges in
a stationary region to constrain this remaining degree of freedom in the BMS 
group.
However, the flux of (super) CM angular momentum is not invariant under
such transformations in a stationary-to-stationary transition, as was 
highlighted in Eq.~\eqref{eq:DeltaKu0}.
To compute this flux, therefore, it is necessary to specify a reference time 
$u_0$ about which it is computed. 
We will use the prescription defined in Eq.~\eqref{eq:u0minDeltaK} that 
minimizes the flux of the CM angular momentum in our computation below.
This then fixes the previously unconstrained time-translation freedom in 
the BMS group.

As was shown in~\cite{Blanchet:2005rj}, through 2PN order, the flux of linear
momentum is parallel to the orbital velocity of the reduced mass of the 
system (and thus the change in the linear momentum is directed radially 
outward).
In terms of the multipole moment $dP^{(\mathrm{GW})}_{1,1}/du$, this is 
related to the fact that the coefficient multiplying $e^{-i\varphi}$ is a 
strictly imaginary quantity (i.e., has vanishing real part) through 2PN order.
With the one real degree of freedom in $u_0$, we can choose this 
reference time to make the flux of the CM part of the angular momentum arising
from the second term in Eq.~\eqref{eq:DeltaK11} vanish through 2PN order.

The first term on the right-hand side of Eq.~\eqref{eq:DeltaK11}, however, 
leads to a change in the CM angular momentum that is $\pi/2$ out of phase with 
that from the second term at 2PN order [i.e., the coefficient multiplying 
$e^{-i\varphi}$ for $dk^{(\mathrm{GW})}_{1,1}/du$ is real].
In addition, the 2.5PN corrections to $dP^{(\mathrm{GW})}_{1,1}/du$ have
terms that are in phase with $dk^{(\mathrm{GW})}_{1,1}/du$.
It is possible to continue canceling the imaginary part of the coefficient
of $e^{-i\varphi}$ of the second term in Eq.~\eqref{eq:DeltaK11} through 2.5PN 
order by appropriately choosing the reference time $u_0$; however, it is not
possible also to cancel the real part of this coefficient in this manner.
This implies that to compute the leading-PN-order expression for the flux of 
the CM angular momentum, we need the leading-order expression for 
$dk_{1,1}^{(\mathrm{GW})}/du$ in Eq.~\eqref{eq:dk11du}, the leading-order 
expression for the time to coalescence in Eq.~\eqref{eq:uofx}, and a 
2.5PN order correction to the leading expression for 
$dP_{1,1}^{(\mathrm{GW})}/du$ in Eq.~\eqref{eq:dP11du} [specifically, the 
part that is in phase with $dk_{1,1}^{(\mathrm{GW})}/du$].
Thus, with this choice of reference time, the two terms on the right-hand side
of Eq.~\eqref{eq:DeltaK11} contribute at the same PN order.

The relevant 2.5PN corrections to the linear momentum flux have been computed
in~\cite{Mishra:2011qz} for nonspinning, quasicircular compact binaries.
We express their result in terms of the $l=1$, $m=1$ moment of the flux by
using the fact that
\begin{equation}
\frac{dP_{1,1}}{du} = -\frac 12 \sqrt{\frac 3{2\pi}} \left( \frac{dP_x}{du}
- i \frac{dP_y}{du} \right) \, ,
\end{equation}
which can be obtained by inverting a relation like the one given in 
Eq.~\eqref{eq:KModesToCartesian}.
It then follows from the results of~\cite{Mishra:2011qz} that 
\begin{align}
\frac{dP^{(2.5\mathrm{PN})}_{1,1}}{du} = & i x^{5/2} \left(
\frac{dP_{1,1}^{(\mathrm{GW})}}{du} \right)
\left( p_{(0)} + p_{(1)} \eta \right) \, ,
\label{eq:dPdu2p5PN}
\end{align}
where we have defined the coefficients
\begin{subequations}
\begin{align}
p_{(0)} = & -\frac{106187}{50460} + \frac{32835}{841}\log 2 - 
\frac{77625}{3364}\log 3 \, ,\\
p_{(1)} = & \frac{10126}{4205} - \frac{109740}{841}\log 2 + 
\frac{66645}{841}\log 3 \, .
\end{align}
\end{subequations}
We can then integrate the flux with respect to $u$.
To evaluate the integral, we change variables to write it as an integral 
with respect to $x$ by using Eq.~\eqref{eq:xdot}.
We find that the result can be expressed in terms of the
fluxes in Eqs.~\eqref{eq:dk11du} and~\eqref{eq:dPdu2p5PN} as 
\begin{align}
\Delta K_{1,1}^{(\mathrm{GW})} = & iM x^{-3/2} \left(
\frac{dk_{1,1}^{(\mathrm{GW})}}{du} + \frac{5M}{256\eta} x^{-4}
\frac{dP_{1,1}^{(2.5\mathrm{PN})}}{du} \right) \nonumber \\
& + O(c^{-2})\, .
\label{eq:CMshiftPN}
\end{align}
The quantity $\Delta K_{1,1}^{(\mathrm{GW})}$ scales with the PN parameter 
$x$ as $x^{5/2} e^{-i\varphi}$.
We have not seen an expression for the change in the CM part of the angular
momentum before, although it is may be related to a part of the 
time-dependent mass-dipole moment computed in~\cite{Blanchet:1993ng}, for
example.

It is relatively straightforward to understand the physics underlying
Eq.~\eqref{eq:CMshiftPN}.
For nonspinning compact binaries in quasicircular orbits, the orbital 
velocity is tangent to the circular orbit up to 2PN order.
At 2.5PN order, however, radiation reaction causes the system to inspiral,
thereby producing a small radial velocity.
It is not possible to remove the effects of this radial velocity on the CM 
angular momentum while preserving the properties of the canonical frame 
associated with the initial stationary region.
This implies that there is a change in the CM part of the angular momentum, 
given by Eq.~\eqref{eq:CMshiftPN}.

To conclude this subsection, we briefly discuss radiation reaction and 
balance equations in PN theory (in the sense of~\cite{Blanchet:1996vx}).
The fluxes of energy and intrinsic angular momentum cause the corresponding 
conserved Poincar\'e charges in the near zone to change at 2.5PN order, and 
the flux of linear momentum also causes such a change in the corresponding 
charge, though at higher (3.5PN) order.
In general, the flux of CM angular momentum also produces a change in the
near-zone CM angular momentum that begins at 3.5PN order.
For nonspinning, quasicircular compact binary sources, however, we showed that 
through an appropriate choice of reference time, the flux of CM angular 
momentum begins at 2.5PN orders higher than the leading effect (which, 
therefore, corresponds to a 6PN-order effect in the near zone).
Because the conserved quantities for nonspinning compact binaries in PN theory 
are currently computed to 4PN order~\cite{Bernard:2017ktp}, this flux leads
to a change in the CM angular momentum that is two PN orders higher than 
the accuracy of the CM angular momentum computed in~\cite{Bernard:2017ktp}.
Thus, we do not anticipate that the flux of CM angular momentum will have a 
significant impact on computations of the dynamics of nonspinning, 
quasicircular compact binaries in the PN approximation.

\subsection{Center-of-mass GW memory effect}

In the two parts of this section, we compute first the null and then the 
ordinary parts of the CM memory effect for nonspinning, quasicircular compact 
binaries in the PN approximation.

\subsubsection{Nonlinear and null part of the CM memory}

We compute the nonlinear part of the null CM memory from the multipolar 
expressions given in Eqs.~\eqref{eq:DeltaKlm} and~\eqref{eq:dkCMlmdu} and the 
relevant definitions of the coefficients that appear in the latter equation.
There are five (independent) nonzero spherical-harmonic modes of this null
nonlinear CM memory at leading PN order, which are given by
\begin{subequations}
\begin{align}
\Delta\mathcal K_{3,1} = & \frac{1}{2880\sqrt{\pi}} \int_{u_1}^{u_2} \! du
[\sqrt{10} (U_{3,3} \dot U_{2,-2} - \dot U_{3,3} U_{2,-2}) 
\nonumber \\
& + 2 \sqrt{6} (U_{3,-1} \dot U_{2,2} - \dot U_{3,-1} U_{2,2}) 
+ 3 U_{2,0} \dot U_{3,1} ] \nonumber \\
& + O(c^{-2}) \, , \\
\Delta\mathcal K_{3,3} = & \frac{1}{2880\sqrt{\pi}} \int_{u_1}^{u_2} \! du
[\sqrt{10} (U_{3,1} \dot U_{2,2} - \dot U_{3,1} U_{2,2}) \nonumber \\
& - 5 U_{2,0} \dot U_{3,3}] + O(c^{-2}) \, , 
\end{align}
for the $l=3$ modes and 
\begin{align}
\Delta\mathcal K_{5,1} = & \frac{1}{50400\sqrt{77\pi}} \int_{u_1}^{u_2} \! du
[(U_{3,3} \dot U_{2,-2} - \dot U_{3,3} U_{2,-2}) +
\nonumber \\
& \sqrt{15} (U_{3,-1} \dot U_{2,2} - \dot U_{3,-1} U_{2,2}) 
- 3 \sqrt{10} U_{2,0} \dot U_{3,1} ] \nonumber \\
& + O(c^{-2}) \, , \\
\Delta\mathcal K_{5,3} = & \frac{1}{50400\sqrt{11\pi}} \int_{u_1}^{u_2} \! du
[\sqrt{10} (U_{3,1} \dot U_{2,2} - \dot U_{3,1} U_{2,2}) \nonumber \\
& - 2U_{2,0} \dot U_{3,3}] + O(c^{-2}) \, , \\
\Delta\mathcal K_{5,5} = & \frac{1}{1680\sqrt{330\pi}} \int_{u_1}^{u_2} \! du
(U_{3,3} \dot U_{2,2} - \dot U_{3,3} U_{2,2} ) \nonumber \\
& + O(c^{-2}) \, ,
\end{align}
\label{eq:DeltaK35modes}%
\end{subequations}
for the $l=5$ modes.
We can then substitute the expressions for the multipole moments in 
Eqs.~\eqref{eq:UlmPNvals} and~\eqref{eq:UlmDotPNvals} to find that
\begin{subequations}
\begin{align}
\Delta\mathcal K_{3,1} = & i\frac{6463}{12600}\sqrt{\frac{\pi}{21}} M \delta m
\eta^2 x^{5/2} e^{-i\varphi}|^{x_2}_{x_1} + O(c^{-2}) \, , \\ 
\Delta\mathcal K_{3,3} = & -i \frac{647}{22680}\sqrt{\frac{\pi}{35}} M \delta m
\eta^2 x^{5/2} e^{-3i\varphi}|^{x_2}_{x_1} + O(c^{-2})\, , 
\end{align}
for the $l=3$ modes and 
\begin{align}
\Delta\mathcal K_{5,1} = & i\frac{677}{154350}\sqrt{\frac{\pi}{330}} M \delta m
\eta^2 x^{5/2} e^{-i\varphi}|^{x_2}_{x_1} + O(c^{-2}) \, , \\
\Delta\mathcal K_{5,3} = & -i\frac{11}{198450}\sqrt{\frac{11\pi}{35}} M 
\delta m \eta^2 x^{5/2} e^{-3i\varphi}|^{x_2}_{x_1} + O(c^{-2}) \, , \\
\Delta\mathcal K_{5,5} = & i\frac{1}{875}\sqrt{\frac{\pi}{77}} M \delta m
\eta^2 x^{5/2} e^{-5i\varphi}|^{x_2}_{x_1} + O(c^{-2}) \, , 
\end{align}
\label{eq:DeltaK35x}%
\end{subequations}
for the $l=5$ modes.
We have used the notation $x_2$ and $x_1$ to denote the values of the
PN parameter at retarded times $u_2$ and $u_1$, 
respectively.\footnote{\label{fn:stationary}~Because $x_1$ and $x_2$ are 
related to the orbital frequency of the binary at times $u_1$ and $u_2$, 
respectively, it is clear that the spacetime is not stationary at either time 
(which breaks one of the assumptions we made in deriving the CM memory effect).
Thus, the results presented in Eq.~\eqref{eq:DeltaK35x} should be taken as
suggestive of how the CM memory effect would grow with $x$, in the PN context
(namely, that it grows in amplitude like $x^{5/2}$, like the change in the CM 
angular momentum does).
The full effect will depend on the details of the merger of the
compact binary and would need to be computed by numerical relativity
simulations of merging compact objects.}
Note that unlike the leading-PN part of the GW memory or the spin memory 
effects, the leading nonlinear, null part of the CM memory effect appears in 
the $m\neq 0$ modes of the multipolar expansion of the effect (specifically
modes with odd $m$ and $l$).
While there are higher-order PN corrections to the GW and the spin memory
effects that appear in the modes with nonzero $m$, it is a distinctive 
feature of the CM memory that the leading-order nonlinear, null CM memory 
effect appears in modes with nonzero $m$.
However, it is also not too surprising, because the flux of CM angular 
momentum for nonspinning, quasicircular compact binaries has no $m=0$ mode 
(only $m=\pm 1$ modes).

\subsubsection{Ordinary part of the CM memory}

A second interesting difference between the GW memory and spin memory effects 
and the CM memory effect is the role of the ordinary part of the memory.
In the PN approximation for nonspinning, quasicircular compact binaries, 
the nonlinear null GW memory appears at leading Newtonian order in the 
waveform, whereas the ordinary part of the memory is typically ignored, 
because it will appear at a PN order that is much higher than that at 
which the PN-expanded gravitational waveform has been computed.
For the spin memory, the ordinary part of the memory is again of a very 
high PN order.

Let us now consider the ordinary part of the CM memory effect.
It was shown in~\cite{Flanagan:2015pxa} that the change in the 
super-CM charges is nonzero when there is GW memory.
To linear order in the GW memory, this change is given by
\begin{equation}
\Delta K_{lm} = -\frac{3M}{16\pi} \sqrt{l(l+1)} \Delta \Phi_{lm} \, ,
\label{eq:KlmChargePhilm}
\end{equation}
where $\Delta \Phi_{lm}$ are the moments of the scalar function $\Delta\Phi$
in Eq.~\eqref{eq:MemDeltaPhi} with respect to scalar spherical harmonics 
(recall that $\Delta K_{lm}$ was computed with respect to 
electric-parity vector spherical harmonics).
The leading GW memory appears in the $m=0$ modes with $l=2$ and $l=4$, and
the values of the potential $\Delta \Phi_{lm}$ are given, for example, 
in~\cite{Nichols:2017rqr}.
Combining the results of~\cite{Nichols:2017rqr} with the expressions in 
Eqs.~\eqref{eq:DeltaKlm} and~\eqref{eq:KlmChargePhilm}, we can then compute 
the leading-PN-order prediction for the ordinary part of the CM memory.
The result is
\begin{subequations}
\begin{align}
\Delta \mathcal K_{2,0} = & -\frac{M}{168}\sqrt{\frac 5\pi} 
\int_{u_1}^{u_2} \! du |\dot U_{2,2}|^2 +O(c^{-2}) \, ,\\
\Delta \mathcal K_{4,0} = & -\frac{M}{453600 \sqrt \pi} 
\int_{u_1}^{u_2} \! du |\dot U_{2,2}|^2 +O(c^{-2}) \, .
\end{align}
\label{eq:DeltaK24modes}%
\end{subequations}
We can then use Eqs.~\eqref{eq:xdot} and~\eqref{eq:UlmDotPNvals} to show that
in terms of the PN parameter $x$, Eq.~\eqref{eq:DeltaK24modes} can be expressed
as
\begin{subequations}
\begin{align}
\Delta \mathcal K_{2,0} = & -\frac{\sqrt{5\pi}}{21}M^2 \eta (x_2-x_1) 
+O(c^{-2}) \, ,\\
\Delta \mathcal K_{4,0} = & -\frac{\sqrt{\pi}}{56700}M^2 \eta (x_2-x_1) 
+O(c^{-2}) \, .
\end{align}
\label{eq:DeltaK24x}%
\end{subequations}
Thus, the nonlinear null part of the CM memory enters at 1.5 PN orders higher 
than the ordinary part of the CM memory.
Moreover, the ordinary part of the CM memory is nonoscillatory ($m=0$) at
leading order, whereas the null part is oscillatory ($m\neq 0$).

The reader might then wonder why we compute the nonlinear null part of
the CM memory, when it is weaker than the ordinary part, for quasicircular,
nonspinning compact binaries.
We do so because, for the CM memory, it will be useful to understand which
terms in the gravitational waveform are responsible for generating the effect.
For the spin memory, there is an easily identifiable term in the GW strain
that produces the effect, when it is integrated in time.
It is also possible, in principle, to measure the terms in the GWs that produce
the spin memory effect with the next generation of ground-based 
interferometers~\cite{Nichols:2017rqr} (and likely space-based interferometers,
too).
To see if the terms in the GWs responsible for the CM memory effect might also
be measured, we must first identify the pertinent terms.
The nonlinear, null part of the CM memory turns out to be the leading PN-order
effect in the GW strain, as we discuss in more detail in the next subsection.

\subsection{GW modes that produce the CM memory effect}

Because the CM memory observable $\Delta\mathcal K$ is a potential for a 
portion of the time integral of the electric-parity part of the GW strain 
(with the terms that grow linearly with $u$ in nonradiative regions removed), 
then there must be terms in the GWs that, when integrated in time, give rise 
to the CM memory effect.
Because the moments $\Delta\mathcal K_{lm}$ are the spherical harmonic modes
of the potential $\Delta\mathcal K$ expanded in scalar harmonics, 
whereas the radiative multipoles $U_{lm}$ correspond to an expansion of the 
GW strain in second-rank, symmetric-trace-free tensor harmonics, there is the
following relationship between these quantities:
\begin{equation}
U_{lm}^{(\mathrm{CM})} = \frac 1{\sqrt 2} \sqrt{\frac{(l+2)!}{(l-2)!}} 
\dot{\mathcal K}_{lm} \, .
\label{eq:UlmToKlmdot}
\end{equation}
We used the notation $U_{lm}^{(\mathrm{CM})}$ to denote just the part of 
$U_{lm}$ that is related to the CM memory ($U_{lm}$ will generally have
other contributions) and $\dot{\mathcal K}_{lm}$ to denote the quantity that
when integrated in time gives rise to $\Delta \mathcal K_{lm}$.

\subsubsection{Nonlinear and null part of the CM memory}

We find that the nonlinear, null part of the CM memory is a consequence
of terms in the gravitational waveform of the form
\begin{subequations}
\begin{align}
U_{3,1}^{(\mathrm{CM})} = & \frac{1}{96\sqrt{30\pi}} 
[2\sqrt{5} (U_{3,3} \dot U_{2,-2} - \dot U_{3,3} U_{2,-2}) 
\nonumber \\
& + 4 \sqrt{3} (U_{3,-1} \dot U_{2,2} - \dot U_{3,-1} U_{2,2}) 
+ 3 \sqrt{2} U_{2,0} \dot U_{3,1} ] \nonumber \\
& + O(c^{-2}) \, , \\
U_{3,3}^{(\mathrm{CM})} = & \frac{1}{96\sqrt{30\pi}} 
[2\sqrt{5} (U_{3,1} \dot U_{2,2} - \dot U_{3,1} U_{2,2}) \nonumber \\
& - 5 \sqrt{2} U_{2,0} \dot U_{3,3}] + O(c^{-2}) \, , 
\end{align}
for the $l=3$ modes and 
\begin{align}
U_{5,1}^{(\mathrm{CM})} = & \frac{1}{1680\sqrt{165\pi}} 
[(U_{3,3} \dot U_{2,-2} - \dot U_{3,3} U_{2,-2}) +
\nonumber \\
& \sqrt{15} (U_{3,-1} \dot U_{2,2} - \dot U_{3,-1} U_{2,2}) 
- 3 \sqrt{10} U_{2,0} \dot U_{3,1} ] \nonumber \\
& + O(c^{-2}) \, , \\
U_{5,3}^{(\mathrm{CM})} = & \frac{1}{240\sqrt{1155\pi}} 
[\sqrt{10} (U_{3,1} \dot U_{2,2} - \dot U_{3,1} U_{2,2}) \nonumber \\
& - 2U_{2,0} \dot U_{3,3}] + O(c^{-2}) \, , \\
U_{5,5}^{(\mathrm{CM})} = & \frac{1}{120\sqrt{154\pi}}
(U_{3,3} \dot U_{2,2} - \dot U_{3,3} U_{2,2} ) + O(c^{-2}) \, ,
\end{align}
\label{eq:U35CMmodes}%
\end{subequations}
for the $l=5$ modes.

Instead of directly substituting the expressions for the multipole moments 
given in Eqs.~\eqref{eq:UlmPNvals} and~\eqref{eq:UlmDotPNvals} into 
Eq.~\eqref{eq:U35CMmodes} to compute the analog of Eq.~\eqref{eq:DeltaK35x}
for the quantities $U_{lm}^{(\mathrm{CM})}$, we note that
for the $m\neq 0$ modes, there is the simple relationship
\begin{equation}
\dot{\mathcal K}_{lm} = -i \frac m M x^{3/2} \Delta \mathcal K_{lm} 
\label{eq:DotDeltaKtoK}
\end{equation}
at the PN order at which we are calculating.
By combining Eqs.~\eqref{eq:DeltaK35x},~\eqref{eq:UlmToKlmdot}, 
and~\eqref{eq:DotDeltaKtoK}, we can easily determine the results for
$U_{lm}^{(\mathrm{CM})}$ in terms of $x$.
It follows that all the moments scale as $x^4 e^{-im\varphi}$, which means
that they are 3PN contributions to the gravitational waveform
(for the $l=3$ modes they are relative 2.5PN-order corrections, and for the
$l=5$ modes, they are relative 1.5PN-order corrections).
Because the 3PN waveform from compact binaries has been computed to this
order~\cite{Blanchet:2008je}, it is possible to compare the expressions for 
$U_{lm}^{(\mathrm{CM})}$ with the equivalent modes in the PN waveform.
There are a few subtleties about making this comparison that we will discuss
further after computing the terms in the GWs that produce the ordinary part 
of the CM memory effect.

\subsubsection{Ordinary part of the CM memory}

Using Eq.~\eqref{eq:UlmToKlmdot} to convert the expressions for 
$\Delta \mathcal K_{l,0}$ in Eq.~\eqref{eq:DeltaK24modes} into expressions
for $U_{l,0}^{(\mathrm{CM})}$, we find that
\begin{subequations}
\begin{align}
U_{2,0}^{(\mathrm{CM})} = & -\frac{M}{84}\sqrt{\frac{15}\pi} |\dot U_{2,2}|^2 
+O(c^{-2}) \, ,\\
U_{4,0}^{(\mathrm{CM})} = & -\frac{M}{75600}\sqrt{\frac 5\pi} |\dot U_{2,2}|^2 
+O(c^{-2}) \, .
\end{align}
\end{subequations}
For these $m=0$ modes, they can be expressed in terms of $x$ as
\begin{subequations}
\begin{align}
U_{2,0}^{(\mathrm{CM})} = & -\frac{128}{7}\sqrt{\frac\pi{15}} M \eta^2 x^5
+O(c^{-2}) \, ,\\
U_{4,0}^{(\mathrm{CM})} = & -\frac{32}{4725}\sqrt{\frac \pi 5} M \eta^2 x^5 
+O(c^{-2}) \, .
\end{align}
\end{subequations}
Because the $U_{l,0}^{(\mathrm{CM})}$ modes scale as $x^5$, then they are
a 4PN effect in the gravitational waveform.
The gravitational waveform at 4PN order has not yet been computed, which 
prohibits us from making a comparison with existing PN results.
However, we anticipate that future PN calculations will find evidence for such 
terms.

\subsection{Comparison with existing PN results}

Because the null part of the CM memory arises from a 3PN effect in the 
gravitational waveform, and because the PN waveform has been computed to this 
accuracy, it would be a useful consistency check of the CM memory effect to 
identify certain terms in the PN expansion of the gravitational waveform 
that are responsible for the CM memory effect.
We find that we can make such an identification in a certain approximation, 
which we will describe in more detail.

Before we do so, however, we must clarify a few notational differences 
between the PN results given in, e.g.,~\cite{Blanchet:2013haa} and those 
in this paper.
The expressions for the PN radiative (as well as canonical and source) 
multipole moments in~\cite{Blanchet:2013haa} are expressed in terms of 
symmetric-trace-free, spatial, rank-$l$ tensors $\mathcal U_L$ and 
$\mathcal V_L$ rather than the multipole moments $U_{lm}$ and $V_{lm}$ (which 
are scalar functions of $u$).
There are well-known prescriptions for converting between the two types of
moments, which are described in~\cite{Thorne1980} (or more recently 
in~\cite{Favata:2008yd}, for example).
The relationships for the radiative mass moments are given by
\begin{subequations}
\begin{align}
U_{lm} = & \frac{16\pi}{(2l+1)!!}\sqrt{\frac{(l+1)(l+2)}{2l(l-1)}} \mathcal U_L
\bar{\mathcal Y}^L_{lm} \, , \\
\mathcal U_L = & \frac{l!}{4} \sqrt{\frac{2l(l-1)}{(l+1)(l+2)}} \sum_m
U^{lm} \mathcal Y_L^{lm} \, ,
\end{align}
\label{eq:ULtoUlm}%
\end{subequations} 
where $\mathcal Y_L^{lm}$ are a set of basis functions for the rank-$l$, 
symmetric-trace-free tensors~\cite{Thorne1980}, and the double factorial means 
a product of all odd integers less than or equal to $(2l+1)$.
Similar relationships exist for the current multipole moments $\mathcal V_L$ 
and $V_{lm}$, though we will not need them in the subsequent discussion.

Having addressed the differences in notation, we must identify the relevant 
terms in the PN waveform.
Because the CM memory effect comes from the integral of the product of 
radiative moments, then the corresponding terms in the PN waveform must be
able to be expressed as an instantaneous product of radiative moments (at
the relevant PN order).
Thus, the other effects at 3PN order in the waveform (contributions from time 
derivatives of 3PN accurate near-zone multipole moments, from components of 
a gauge transformation needed to relate the near-zone moments to the 
intermediate canonical moments, and from tail and hereditary terms) will not 
be needed here.
The instantaneous and nonlinear terms in the 3PN-accurate radiative moments, 
however, are expressed in terms of the canonical moments.
We now reproduce the expressions for these parts of the $l=3$ and $l=5$ 
radiative moments, which can be found, for example, in Eqs.~(95a) and~(95e) 
of~\cite{Blanchet:2013haa}:
\begin{subequations}
\begin{align}
\mathcal U_{ijk}^{(\mathrm{IN})} = & 
-\frac 43 M^{(3)}_{a\langle i}M^{(3)}_{jk\rangle a}
-\frac 94 M^{(4)}_{a\langle i}M^{(2)}_{jk\rangle a}
+\frac 14 M^{(2)}_{a\langle i}M^{(4)}_{jk\rangle a} \nonumber \\
& -\frac 34 M^{(5)}_{a\langle i}M^{(1)}_{jk\rangle a}
+\frac 14 M^{(1)}_{a\langle i}M^{(5)}_{jk\rangle a}
+\frac 1{12} M^{(6)}_{a\langle i}M_{jk\rangle a} \nonumber \\
& +\frac 14 M_{a\langle i}M^{(6)}_{jk\rangle a} \, , \\
\mathcal U_{ijkpq}^{(\mathrm{IN})} = & 
-\frac {710}{21} M^{(3)}_{\langle ij}M^{(3)}_{kpq\rangle} 
-\frac {265}{7} M^{(4)}_{\langle ij} M^{(2)}_{kpq\rangle} \nonumber \\
& -\frac{120}{7} M^{(2)}_{\langle ij} M^{(4)}_{kpq\rangle}
-\frac{155}{7} M^{(5)}_{\langle ij} M^{(1)}_{kpq\rangle} - \nonumber \\
& \frac{41}{7} M^{(1)}_{\langle ij} M^{(5)}_{kpq\rangle} 
-\frac{34}{7} M^{(6)}_{\langle ij} M_{kpq\rangle} 
-\frac{14}{7} M_{\langle ij} M^{(6)}_{kpq\rangle} \, .
\end{align}
\label{eq:U35modesBlanchet}%
\end{subequations}
The superscript ``(IN)'' is short for ``instantaneous and nonlinear,''
the repeated index $a$ is being summed over in the first three lines, 
and the angled brackets mean to take the symmetric trace-free part of the
tensor.

It is not immediately obvious how to relate the products of canonical moments 
that appear in Eq.~\eqref{eq:U35modesBlanchet} to the products of radiative
moments that appear in the GW modes that produce the CM memory effect.
The reason is that the canonical moments that appear in 
Eq.~\eqref{eq:U35modesBlanchet} have fewer than $l$ derivatives with respect 
to time (where $l$ is the multipole order of the different canonical moments 
that appear in the products of the moments).
Thus, we cannot directly use the analog of the relationships in 
Eq.~\eqref{eq:RadToCanonical} for the rank-$l$ symmetric-trace-free tensors 
to express the radiative moments in terms of the canonical moments.
Instead, we would have to express these derivatives of the canonical moments 
in terms of integrals of the radiative moments by integrating an expression 
like Eq.~\eqref{eq:RadToCanonical}.
In performing this procedure, we would need to introduce new constants of 
integration, but we do not have a prescription for determining the values
of these constants.

Because the CM memory effect involves a time integral of the GW strain, 
however, it is equally relevant to know whether the time integral of the PN 
expressions in Eq.~\eqref{eq:U35modesBlanchet} agree with the time integral of 
the modes in Eqs.~\eqref{eq:U35CMmodes} [after using the relationships in 
Eq.~\eqref{eq:ULtoUlm}].
In making this comparison, we can integrate by parts
to obtain an equivalent expression that involves a different linear combination
of derivatives of the canonical moments (and boundary terms from integrating
by parts).
If the boundary terms vanish, then the integrand is a new expression for the
relevant parts of the radiative moments that give rise to the same CM memory 
effect.\footnote{A subtle issue will be whether the boundary terms vanish.
This clearly will not be true if we consider the binary as it evolves between
two nonzero frequencies $x_1$ and $x_2$, but this will also break the 
assumption of a stationary-to-stationary transition (as was further discussed
in footnote~\ref{fn:stationary}).
Thus, we will consider the full evolution of the binary as it makes
a stationary-to-stationary transition; this will eliminate the majority of 
the boundary terms.
However, there are some additional boundary terms that will not vanish 
in stationary regions, which occur because of the GW memory effect.
These terms are of a sufficiently high PN order that we will not need to
treat them at the PN accuracy at which we perform the calculation.
As a result, we are able to ignore boundary terms when integrating by parts
when we make this comparison.}

We will use this procedure of integrating in time, integrating by parts, 
and differentiating the expression to get a new PN expression for the 
instantaneous, nonlinear terms.
In this procedure, we will integrate by parts so that we can write the result
in terms of products of the radiative moments and their derivatives (but not
their integrals).
This will avoid issues with unknown constants of integration, which
were mentioned above.
The result of this process is that Eq.~\eqref{eq:U35modesBlanchet} can be 
written as
\begin{subequations}
\begin{align}
\mathcal U_{ijk}^{'(\mathrm{IN})} = & \frac 1{12} 
(\mathcal U_{a\langle i} \dot{\mathcal U}_{jk\rangle a} 
- \dot{\mathcal U}_{a\langle i} \mathcal U_{jk \rangle a} ) \, , \\
\mathcal U_{ijkpq}^{'(\mathrm{IN})} = & \frac{2}{21} 
(\mathcal U_{\langle i j} \dot{\mathcal U}_{kpq\rangle} 
- \dot{\mathcal U}_{\langle ij} \mathcal U_{kpq\rangle} ) \, .
\end{align}
\end{subequations}
We have added an apostrophe to the modes $\mathcal U^{'(\mathrm{IN})}_L$ to
indicate that they were obtained from the expressions for 
$\mathcal U^{(\mathrm{IN})}_L$ in Eq.~\eqref{eq:U35modesBlanchet} by 
integrating by parts and differentiating the resulting expression [as well as
using the analog of Eq.~\eqref{eq:RadToCanonical} for the symmetric-trace-free
tensors].
With the relationships in Eq.~\eqref{eq:ULtoUlm}, we can then recover the
$l=3$ and $l=5$ modes given in Eq.~\eqref{eq:U35CMmodes}, namely,
\begin{equation}
U^{'(\mathrm{IN})}_{3m} = U^{(\mathrm{CM})}_{3m} \, , \qquad
U^{'(\mathrm{IN})}_{5m} = U^{(\mathrm{CM})}_{5m} \, ,
\end{equation}
for odd integers $m$.
Therefore, there are terms in the already computed 3PN waveform that 
give rise to the same CM memory effect, under the prescription described above
for rewriting the instantaneous and nonlinear terms in the 3PN waveform.

\subsection{Discussion of PN results}

Because the CM memory effect arises from 3PN and 4PN terms in the GWs
from a compact binary, it is of interest to determine whether these terms
in the gravitational waveform could be detected by any current or upcoming 
GW observatories.
The GW memory could be detected within the next decade after LIGO (as well
as Virgo and KAGRA) detects hundreds of binary-black-hole 
mergers~\cite{Lasky:2016knh}.
This is possible because the effect enters at leading (Newtonian) order in
the waveform, and it has a distinctive dependence on time and on angular 
coordinates (it is nonoscillatory, and enters into the $m=0$ and $l=2,4$ 
modes of the gravitational waveform for nonspinning, quasicircular compact 
binaries at leading order).
The spin memory effect also has distinctive time and angular dependencies 
(it enters into the $l=3$, $m=0$ mode of the time-integrated gravitational
waveform for nonspinning, quasicircular compact binaries at leading order); 
however, the related terms in the GWs are of 2.5PN order in the waveform.
This means that it will likely be too weak to be detected by the current 
generation of ground-based detectors, but it could conceivably be observed 
by the next generation of ground-based GW detectors, like the Einstein 
Telescope~\cite{Nichols:2017rqr}.

The GW modes related to the CM memory effect seem much more difficult to 
detect.
For the nonlinear part, the modes appear as a 3PN order term in the 
waveform.
Specifically, for the $l=3$ modes of nonspinning, quasicircular compact
binaries, they are a 2.5PN-order correction to GW 
modes that vanish when the components of the binary have the same mass (and 
thus are themselves a correction to the leading quadrupole waveform).
While the small amplitude of the effect will make detecting it challenging, 
there are two other properties of the PN waveform that seem to prohibit 
being able to identify the terms in the GWs that produce the nonlinear null 
part of the CM memory effect.
First, although we showed that $U_{lm}^{(\mathrm{IN})}$ can be reexpressed as 
$U^{'(\mathrm{IN})}_{lm}$ [or equivalently $U_{lm}^{(\mathrm{CM})}$] for 
harmonics with $l=3,5$ and $m$ odd in a stationary-to-stationary transition,
outside of this context, $U_{lm}^{(\mathrm{IN})}$ and $U_{lm}^{(\mathrm{CM})}$
can be different.
Second, at 3PN order, there are additional terms that arise from nonlinear 
interactions in the near zone of the compact binary that produce effects in 
the gravitational waveform that have (at least at this PN order) the same time 
dependence as those responsible for the CM memory (but they would have a
different dependence on angular coordinates).
The full gravitational waveform is a sum of these different contributions, and 
it is not clear how observationally to separate out the part related to the 
nonlinear null CM memory effect from these other similar effects from a given
compact-binary source.

Next, we consider the ordinary part of the CM memory effect.
For nonspinning, quasicircular compact binaries, it is a 4PN correction to 
the same GW multipole moments in which the GW memory appears.
While it is of a high PN order, it has a different angular dependence 
(the ratio of the $l=2$ and $l=4$ modes differs from that of the GW memory).
Perhaps more importantly, it also has a different time dependence than the 
GW memory does.
It grows with time like the instantaneous flux of energy does, unlike 
the GW memory, which grows with time as the total radiated energy does.

We can roughly estimate whether these modes are detectable by computing the 
signal-to-noise ratio of the part of the GWs that produce the ordinary CM 
memory effect.
For our source, we choose a binary, like the first GW150914 detection by 
LIGO~\cite{Abbott:2016blz}, and for our detector, we use the Einstein 
Telescope (specifically the analytical fit for the ET-B noise curve given 
in~\cite{Regimbau:2012ir}).
An event like GW150914 will likely be one of the loudest events to be observed 
by the Einstein Telescope, because its signal-to-noise ratio could be in the 
thousands~\cite{Nichols:2017rqr}.
Following a procedure similar to that described in~\cite{Nichols:2017rqr} to 
compute the signal-to-noise ratio, we find that the GW modes that produce the
ordinary part of the CM memory effect have a signal-to-noise ratio that is 
several orders of magnitude less than unity.
Thus, it is difficult to imagine that it will be detected by ground-based
GW detectors from individual events.
Attempting to stack multiple events to build evidence for the CM memory also 
seems difficult, because the amplitude of the effect in the GWs is 
significantly smaller than the background noise in the detector.
The prospects for other detectors like the space-based LISA 
mission~\cite{Audley:2017drz} or pulsar timing arrays 
(e.g.,~\cite{IPTA:2013lea}) we expect will be similar.

\section{\label{sec:Conclusions} Conclusions}

In this paper, we investigated the flux of (super) angular momentum in 
asymptotically flat spacetimes.
We showed that within the context of stationary-to-stationary transitions
the change in the (super) angular momentum between two cuts is not invariant
under supertranslations.
The difference produced by a supertranslation is related to the change in 
supermomentum, the GW memory, and the supertranslation itself.
Next, we focused on the flux of the center-of-mass part of the
angular momentum. 
We argued that the change in the (super-) CM angular momentum (although not 
invariant under supertranslations) contains additional information about an 
isolated system that is not contained in the change in the 4-momentum, 
intrinsic (super) angular momentum, supermomentum, GW memory, or spin memory.
We then derived a new multipolar expression for the flux of CM angular momentum
in terms of a set of radiative multipole moments of the GW strain.

The next part of the paper was devoted to defining the CM memory effect.
The effect is related to the time integral of the electric-parity part of the
GW strain, with the part that grows linearly with retarded time (from the 
ordinary GW memory) removed.
The quantity we defined is invariant under infinitesimal supertranslations.
We then derived an expression for the multipole moments of this CM memory 
effect in terms of the radiative multipoles of the GW strain and the multipole
moments of the change in the super-CM angular momentum.

The final part of the paper was devoted to analyzing nonspinning, 
quasicircular compact binaries, which we treated in the post-Newtonian 
approximation.
We showed that binaries with components with unequal masses will typically
have a nonzero flux of CM angular momentum.
The effect was quite weak (of a high PN order), because with
the freedom to shift the reference time about which the flux is computed,
it was possible to set the change in the CM angular momentum to be zero
through 2.5PN order (which corresponds to a 6PN-order effect in the 
near-zone equations of motion).

Lastly, we computed the CM memory effect for these binaries, and we found that
the ordinary part of the CM memory was a larger (lower PN-order) effect than
the nonlinear null part of the memory.
The opposite is true for the GW memory and the spin memory effects.
The nonlinear, null part of the CM memory arises from a 3PN term in the GWs, 
which we could identify with a certain part of the 3PN gravitational
waveform from nonspinning, quasicircular compact binaries.
The ordinary part of the CM memory comes from a 4PN term in the GWs, which 
has not yet been computed in PN theory.
The null part of the CM memory effect turned out to be degenerate with other 
nonlinear terms in the PN waveform, which made it seem difficult to 
identify and measure the effect with current or future GW detectors.
The ordinary part of the CM memory is measurable in principle, but it was 
sufficiently weak that it seemed unlikely that any upcoming GW interferometers 
or a pulsar timing array would be able to observe the effect.
Thus, we suspect that the results in this paper will be more pertinent for
helping to understand the theoretical properties of the extended BMS charges 
than for highlighting observable GW effects related to the changes in these 
charges from compact binaries.

\acknowledgments

It is my pleasure to thank B\'eatrice Bonga, Yanbei Chen, Zachary Mark, 
Leo Stein, and Aaron Zimmerman for helpful discussions about 
the properties of the center-of-mass part of the angular momentum.
I am also grateful to Thomas M\"adler for his correspondence about the 
evolution equations in the Bondi-Sachs framework and his suggestions for 
improvements in Sec.~\ref{sec:CMflux}.
I am appreciative for the useful comments Samaya Nissanke provided on a 
draft of this paper.
Finally, I thank an anonymous referee for pointing out several important 
references that I had overlooked.
This work is part of the research program Innovational Research Incentives
Scheme (Vernieuwingsimpuls), which is financed by the Netherlands Organization
for Scientific Research through the NWO VIDI Grant No.~639.042.612-Nissanke.

\bibliography{Refs}

\end{document}